\documentclass{article}

\usepackage{subcaption}
\usepackage{PRIMEarxiv}
\usepackage{amsmath}
\usepackage{comment}
\usepackage[utf8]{inputenc} 
\usepackage[T1]{fontenc}    
\usepackage{hyperref}       
\usepackage{url}            
\usepackage{booktabs}       
\usepackage{amsfonts}       
\usepackage{nicefrac}       
\usepackage{microtype}      
\usepackage{lipsum}
\usepackage{fancyhdr}       
\usepackage{graphicx}       
\graphicspath{{./img/}}     

\usepackage{color}

\newcommand{\ul}[1]{\textbf{ {\textcolor{blue}{#1 -- UL}}}}

\usepackage{makecell}
\usepackage{float}

\title{DCAE-SR: design of a Denoising Convolutional Autoencoder for reconstructing Electrocardiograms signals at Super Resolution.}

\author{
    Ugo Lomoio \\
  Department of Surgical and Medical Sciences \\
  Magna Graecia University of Catanzaro \\
  Catanzaro,  Italy\\
  \texttt{ugo.lomoio@unicz.it} \\
  \And 
    Pierangelo Veltri \\
  DIMES \\
  University of Calabria \\
  Rende, Italy\\
  \texttt{pierangelo.veltri@dimes.unical.it} \\
  \And 
     Pietro Hiram Guzzi \\
  Department of Surgical and Medical Sciences \\
  Magna Graecia University of Catanzaro \\
  Catanzaro, Italy\\
  \texttt{hguzzi@unicz.it} \\
  \And
    Pietro Liò \\
  Department of Computer Science and Technology \\
  Cambridge University \\
  Cambridge, United Kingdom\\
  \texttt{pl219@cam.ac.uk}  \\
}

\begin{document}
\maketitle
\begin{abstract}

Electrocardiogram (ECG) signals play a pivotal role in cardiovascular diagnostics, providing essential information on the electrical activity of the heart. However, the inherent noise and limited resolution in ECG recordings can hinder accurate interpretation and diagnosis. In this paper, we propose a novel model for ECG super resolution (SR) that uses a DNAE to enhance temporal and frequency information inside ECG signals. Our approach addresses the limitations of traditional ECG signal processing techniques.  Our model takes in input 5-second length ECG windows sampled at 50 Hz (very low resolution) and it is able to reconstruct a denoised super-resolution signal with an x10 upsampling rate (sampled at 500 Hz). We trained the proposed DCAE-SR on public available  myocardial infraction ECG signals. Our method demonstrates superior performance in reconstructing high-resolution ECG signals from very low-resolution signals with a sampling rate of 50 Hz. We compared our results with the current deep-learning literature approaches for ECG super-resolution and some non-deep learning reproducible methods that can perform both super-resolution and denoising. We obtained current state-of-the-art performances in super-resolution of very low resolution ECG signals frequently corrupted by ECG artifacts. We were able to obtain a signal-to-noise ratio of 12.20 dB (outperforms previous 4.68 dB), mean squared error of 0.0044 (outperforms previous 0.0154) and root mean squared error of 4.86\% (outperforms previous 12.40\%). In conclusion, our DCAE-SR model offers a robust (to artefact presence), versatile and explainable solution to enhance the quality of ECG signals. This advancement holds promise in advancing the field of cardiovascular diagnostics, paving the way for improved patient care and high-quality clinical decisions.




\end{abstract}

\thanks{\textit{\underline{Citation}}: \textbf{Ugo Lomoio, Pierangelo Veltri, Pietro Hiram Guzzi, Pietro Liò. DCAE-SR: design of a Denoising Convolutional Autoencoder for reconstructing ElectroCardioGrams signals at Super Resolution.}}


\section{Introduction}

The physiological activity of the human organs can be investigated by measuring corresponding electrical, chemical, or mechanical signals \cite{prince2006medical} to improve their understand and disease diagnosis and monitoring \cite{meyer2004pattern}. For example, electrical signals from the heart aid in diagnosing cardiac diseases, whereas chemical signals, such as glucose levels in the blood, help manage diabetes \cite{satija2018review}. 

Hearth signals are usually stored into the so called electrocardiograms (ECGs) after the acquisition,  processing, and analysis \cite{wasimuddin2020stages}. Signal acquisition and analysis has promoted the development of personalized medicine, where treatments can be tailored to an individual's specific needs based on the unique signals their body produces. By integrating computational models and real-time data analysis, signals in biomedicine are at the forefront of improving patient outcomes and advancing healthcare practices \cite{siontis2021artificial}.

In this paper we focus on the analysis of ECGs \cite{ajdaraga2017fs}. Each ECG of a patient includes up to twelve signals, also called ECG channels (or leads), corresponding to a specific electrode which capture a particular heart's electrical activity from a different perspective. The signals include six limb leads (I, II, III, aVR, aVL, and aVF) and six precordial leads (from V1 to V6), providing information about the heart's electrical axis. The number of electrodes is related to the specific aim of the analysis: for example for long monitoring of a patients (e.g. days or weeks), a 3-lead ECG is often used; while in clinical practice  12-lead ECG  is commonly used. 

Given the rhythmic nature of the heart, ECG systems must be designed to capture and record only signals related to a certain range of frequencies associated to the heart activity \cite{kligfield2007prevalence}. The standard ECG bandwidth recommended by the American Heart Association (AHA) is defined as the frequency interval that goes from 0.05 Hz to 150 Hz. When using frequencies lower than 0.05 Hz, it is possible that the so called the baseline wander (BW) artifact that leads to slow variations in the ECG baseline and can be caused by patient respiratory activities \cite{garcia2009technical}. Artifacts caused by muscular activities may be corrupt the signals when frequencies greater than 150 Hz are used. 

Signals captured by using ECG instruments are sampled at different frequencies, i.e. the number of samples captured for each second \cite{pizzuti1985digitalsampling}. A higher sampling frequency provides a more detailed representation of the ECG signal, by capturing rapid changes in electrical activity allowing for more accurate detection and analysis of cardiac events \cite{bui2021comparison,shekatkar2017detecting,narayanaswamy2002HR,ajdaraga2017fs}. Nowdays, the standard ECG sampling rate recommended by the AHA is 500 Hz.

In many clinical settings, a lower frequency of 50 or 100 Hz (Low Resolution - LR), is used since both are sufficient to capture the minimum required detail in the waveform needed for manual analysis. However, sampling rate of 500 or 1000 Hz (High Resolution - HR) or even more can be required for specific research purposes or for the diagnosis pathologies caused by rapid changes in the electrical activity of the heart such as, for example, identification of late potentials \cite{berbari1998latepot}. Moreover, in some scenarios, HR signals must be downsampled to efficiently transmit over low-bandwidth channels \cite{huang2023ecgsr}.  Applying super-resolution (SR), i.e.  obtaining high-resolution ECG signals from  low-resolution versions, may give more details on the signals, better denoising power and more possibility to detect modification related to diseases \cite{kaniraja2024deep,chen2023srecg}. The advantages of super-resolution (SR) signal analysis span a large field:
\begin{itemize} 
\item signal enhancement for denoising, artifact removal, and the refinement of signal details to identify critical cardiac events \cite{arsene2019deep}; 
\item extraction of detailed features that are not discernible at lower resolutions, such as precise measurement of P, QRS, and T waveforms \cite{wu2020extracting};
\item analysis of Microvolt T-Wave Alternans (TWA), which is a subtle fluctuation in the amplitude of the T wave in ECG signals and is considered a marker of ventricular arrhythmias and sudden cardiac death \cite{burattini2009comparative}; 
\item Arrhythmia Detection and Classification through the increased detail available in super-resolution ECG signals \cite{hannun2019cardiologist}.
\end{itemize}

Despite these advantages, it is not easy to obtain super resolution without the use of computational advantages of deep learning frameworks. For these aims, we  introduces a novel approach for ECG super-resolution based on a Denoised Convolutional AutoEncoder (DCAE) modified to perform both reconstruction, denoising, and super-resolution tasks. 

\begin{figure}[H]
    \centering
    \includegraphics[height=10cm,width=15cm]{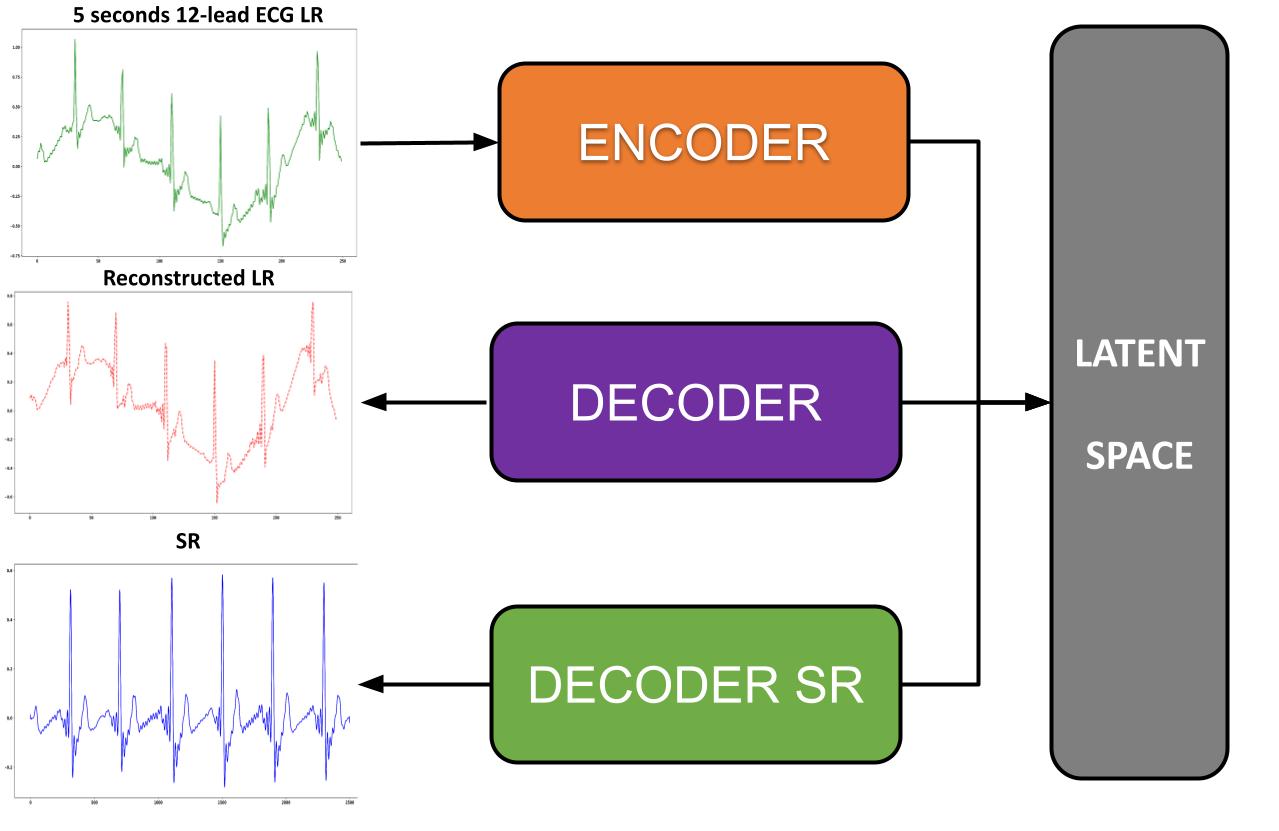}
    \caption{Our proposed DCAE-SR: a modified DCAE Encoder-Decoder architecture with a second Decoder for denoising and super-resolution (the DECODER SR).}
    \label{fig:figArchitectureIntro}
\end{figure}

The architecture of the proposed DCAE-SR, as depicted in \ref{fig:figArchitectureIntro}, is based on the use of a modified Encoder-Decoder architecture in which two decoders are used: one is able to reconstruct signals at the same original frequency, while a second one has been trained to derive a reliable high resolution version of the signals. 

We tested the proposed DCAE-SR on the PTB-XL dataset \cite{ptbxl, physionet}, a large dataset of 21799 clinical 12-lead 10-second multi-annotated ECG signals from 18869 different patients. Our  results obtained by our DCAE-SR model on the PTB-XL dataset show the improvements with respect to state-of-the-art approaches.

\begin{figure}[ht]
    \centering
    \includegraphics[height=6cm, width=16cm]{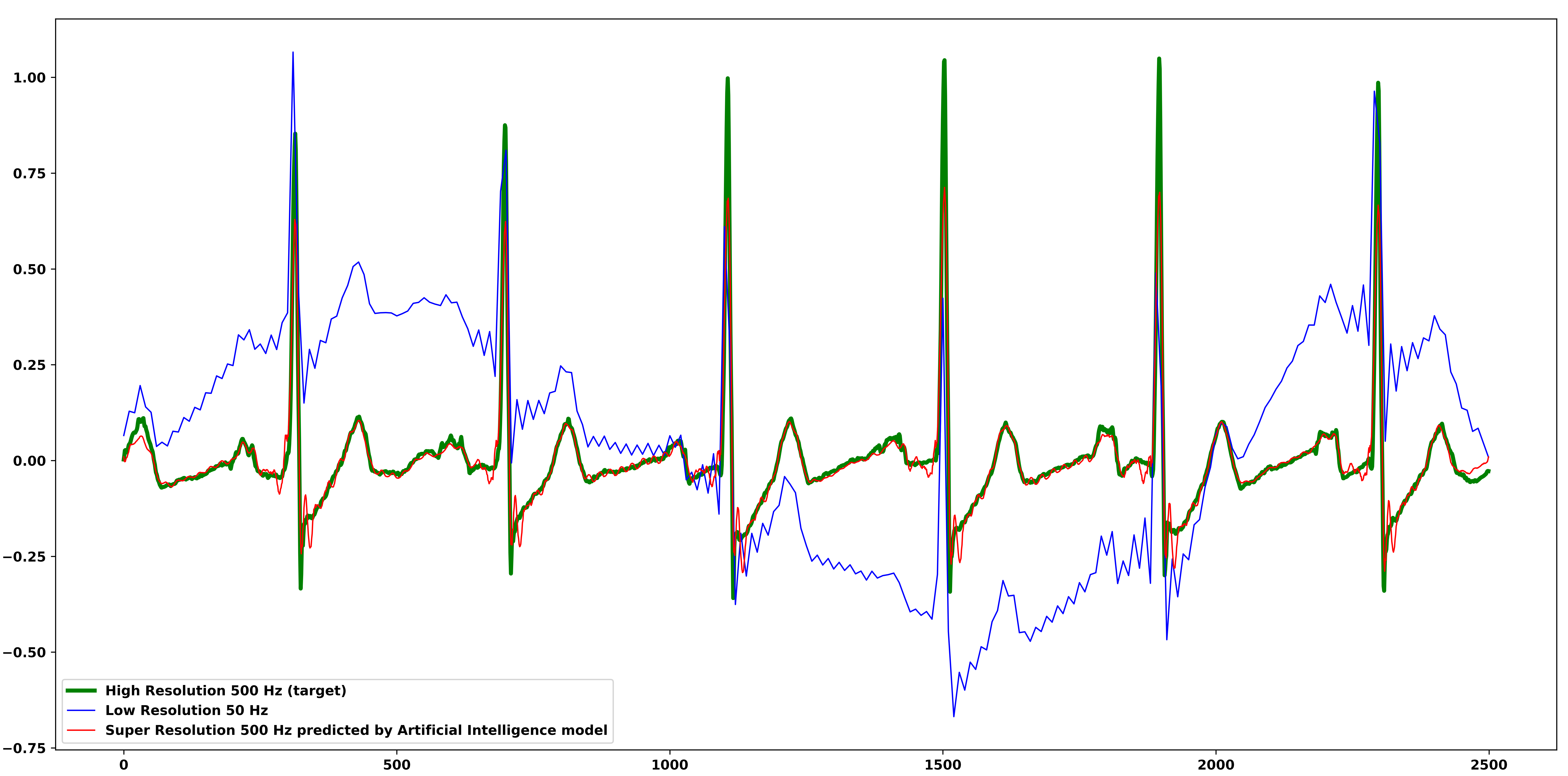}
    \caption{Input: LR corrupted signal sampled at 50 Hz (blue), Target: HR clean signal sampled at 500 Hz (green), Output: predicted SR denoised signal sampled at 500 Hz. Lead I only reported for simplicity.}
    \label{fig:figInputOutput}
\end{figure}

\begin{figure}[H]
    \centering
    \includegraphics[height=10cm, width=16cm]{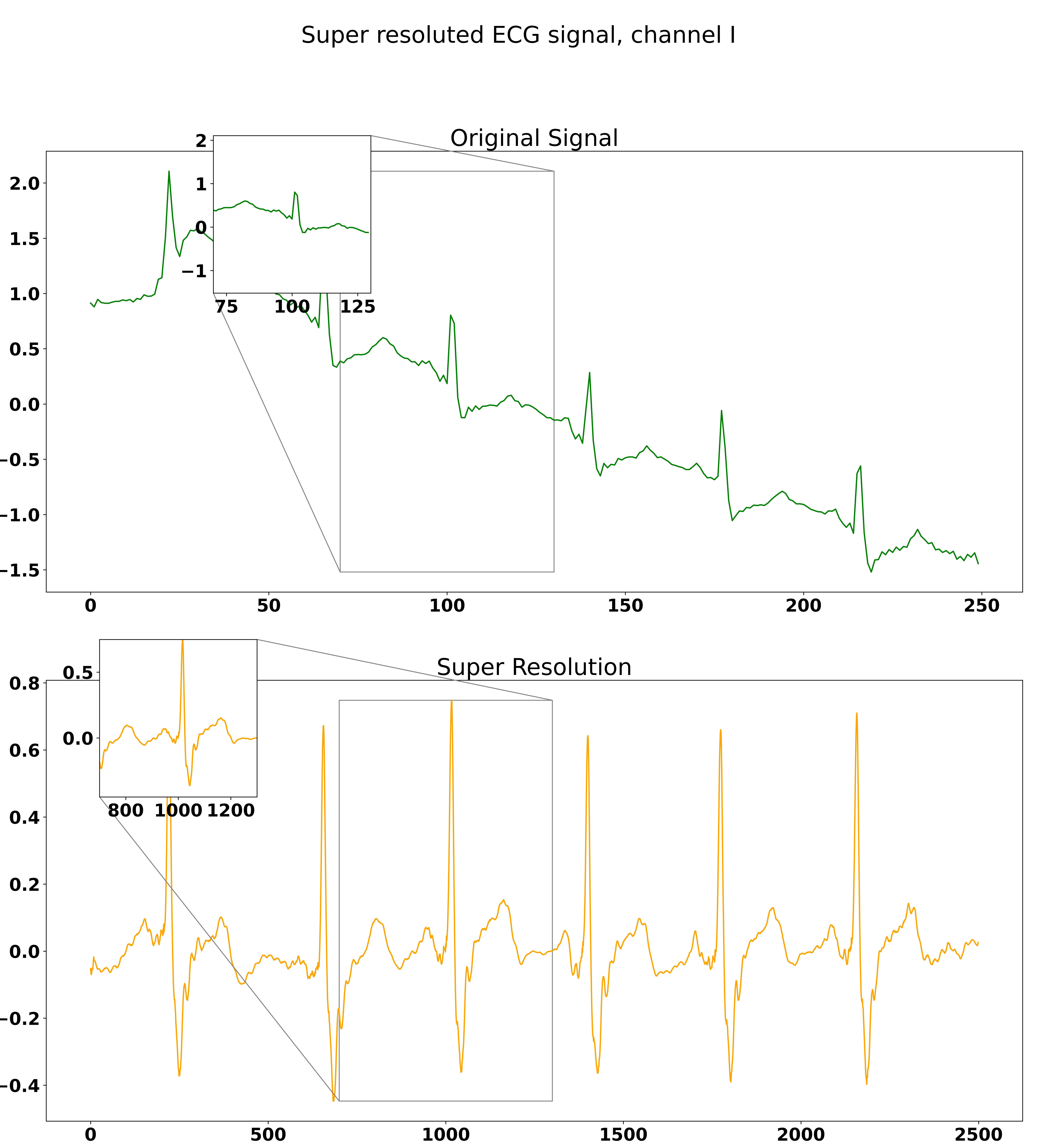}
    \caption{Corrupted LR 50 Hz in input of our DCAE-SR (up), 500 Hz cleaned super-resolution (bottom)}
    \label{fig:figLRSR}
\end{figure}

In Figure \ref{fig:figInputOutput} we report an example of the proposed method. An input lower resolution signal (reported in blue) with noise and artifacts is given as input. The output of the system (represented in red) is the denoised HR signal which is compared with the \textit{true} HR signal (represented in green). Figure \ref{fig:figLRSR} presents a more detailed comparison between a LR signal with noise and the obtained HR.

The principal contributions and innovations of this research can be summarized as follows:

\begin{itemize}
    \item we developed an innovative solution that addresses some of the current limitations in ECG signal acquisition, such as, artifacts presence or very low-resolution sampling rate;
    \item we are the first that take in consideration noise and artifact presence inside very low resolution signals sampled at 50 Hz; 
    \item to the best of our knowledge, we are the first to propose an extension of the DCAE architecture to perform ECG denoising super-resolution task;
    \item we applied model explainability to elucidate how the model identifies and utilizes relevant features or patterns in the LR input to generate the denoised SR output;
    \item we reported a robust behaviour and performances of our model in both noisy and clean LR ECG signals;
\end{itemize}

\section{Related Work}


Autoencoders (AE) \cite{bank2021autoencoders,gu2022modeling}, which serve as the foundational architecture of our super-resolution approach \cite{dong2015imagesr}, excel in the tasks of unsupervised learning by adeptly capturing the intricate structural complexities inherent in ECG signals. Beyond their role in super-resolution, AEs find applications in various fields like input reconstruction (compress and restore the original input), denoising, and anomaly detection (where deviations from learned patterns can be indicative of irregularities in the ECG signal) \cite{lomoio2023ecg}. 

Convolutional Autoencoders (CAEs) introduce convolutional layers inside the AE architecture, significantly augmenting the model's ability to extract spatial and temporal features from signals \cite{zhang2018CAE}, which is particularly indicated in the time-frequency domain of ECG signals, enabling better extraction of features from signals that can result in better input reconstruction, anomaly detection, or input super-resolution.

The presence of noise or artifacts inside ECG signals poses a significant challenge for AEs and CAEs due to their inherent sensitivity. In the case of super-resolution, noise and artifact presence inside the LR input data can adversely affect models' performance, leading to degraded super-resolution quality and compromised feature extraction \cite{fariha2020ecgnoise, perez2018artifacts} caused by augmenting the resolution of both signal and noise. We need a solution that can automatically distinguish between ECG signal features and ECG artifact features inside the latent space; once we have this, we can apply super-resolution only on the ECG signal and simultaneously remove unwanted artifacts.

Incorporating denoising techniques inside a CAE leads to introducing denoising CAE (DCAE) \cite{dcae2008}. DCAEs learn robust representations by training on corrupted versions of the input data and reconstructing the clean and uncorrupted versions. This approach helps the model focus on extracting meaningful features while disregarding noise \cite{dcae2016images,guzzi2022editorial}. Additionally, regularization techniques, such as dropout, can be employed during training to enhance the model's generalization and noise resilience.


The image or audio signal super-resolution field has been extensively explored in recent years \cite{lio2020srmri, lio2019lesionsr, kuleshov2017audio}. The first significant, but not deep learning, super-resolution method is often considered to be the interpolation-based approach, which involves the use of interpolation techniques to increase the spatial resolution of images or signals \cite{castiglioni2003interpolation, Han2013ComparisonOC}. 

Thanks to recent advances in the field of deep learning, together with the increase in availability and performance of modern Graphics Processing Units (GPUs) \cite{gpu2008}, new deep learning SR methods that outperform the basic interpolation methods have come to light. 

Even if modern deep learning techniques outperformed the interpolation approach, some of the first deep learning super-resolution techniques, called pre-upsampling SR methods, incorporate at least one interpolation layer as a pre-processing step that performs a not-learnable upsampling on the LR input and then learns how to reconstruct the HR data from the LR upsampled input \cite{chen2022SRreview, mekhlafi2024SRreview}. We can distinguish deep learning super-resolution techniques based on where the upsampling layer is applied:
\begin{itemize}
    \item pre-upsampling SR methods: In these methods, an upsampling layer, in most cases a not-learnable upsampling such as interpolation, is the first layer of the model architecture (often used as a pre-processing step). An example of a pre-upsampling SR deep learning method is the SR-CNN model \cite{SRCNN}, which applies a bicubic interpolation layer to augment the spatial resolution of the input;
    \item post-upsampling SR: In these methods, the upsampling layer is placed as the last layer of the signal. An example of a post-upsampling SR deep learning method is the FSRCNN \cite{FSRCNN} model, which uses deconvolutional (transposed convolutional) layers for learnable super-resolution. Our SR method, which uses multiple deconvolutional layers to perform LR noised reconstruction and denoised SR, can be considered a post-upsampling SR method with denoising power;
    \item progressive-upsampling SR: In these methods, upsampling layers are progressively placed inside the network architecture \cite{lyu2020progressivesr}.
\end{itemize}

One of the first proposed methods for ECG super-resolution was SRECG \cite{SRECG}, an improvement of the well-known SRResNet \cite{SRResNet}. SRECH aimed to introduce a system based on a super-resolution model to speed up the transmission of ECG signals from a portable/wearable (P/W) device. It was based on the compression of a HR signal into a LR which is transmitted into a low-bandwidth channel and the final reconstruction where an automated classifier was used to find arrhythmia. The SRECG model was trained and evaluated using the China Physiological Signal Challenge 2018 dataset (CPSC2018) \cite{CPSC2018}. The evaluation of the super-resolution is performed only at the classification level; no results report the overall quantitative super-resolution performance metrics (such as mean PSNR, SNR, MSE, or RMSE on the validation set).


Huang et al. \cite{huang2023ecgsr} successively presented a Signal-Referenced Network \cite{cao2022referencebased} for 3-lead (I, II, V2) ECG super-resolution. This Signal-referenced Network for ECG super-resolution takes in input a signal comprised of some leads compressed at low resolution (for example, I, II) and a referenced high-resolution lead (for example, V2), giving in output all three leads in high resolution. For a fair comparison with our and other super-resolution models, we will present only their super-resolution results regarding the 3-lead low-resolution (50 Hz) experiment, without any high-resolution references in the input (group 4, index 10 in their paper). They report a 0.15 mV RMSE between the obtained 500 Hz super-resolution signal and the target 500 Hz high-resolution.

A recent paper in March 2024 by Kaniraja et al.  \cite{SRCNN-ECG} proposed a modified version of the well-known pre-upsampling SRCNN adapted for 1-dimensional signals such as ECGs. The SRCNN-ECG model  takes in input all 12 leads of the ECG signal in low resolution (example, sampled at 50 Hz) and it preprocesses the input  using a bicubic interpolation layer, with a given sampling rate (10, for example), as a pre-upsampling super-resolution to augment the temporal resolution of the signal. Then it uses a 5-layer Convolutional RESidual NETwork (ResNet) \cite{resnet} that computes the super-resolution. Finally, it  gives as output the super-resoluted signal sampled at 500 Hz.  The model was trained and evaluated on a 1500-signals random subset of the CPSC2018 dataset. They report the following super-resolution performances on the validation set using an upscaling rate of 10 from LR 50 Hz to HR 500 Hz: RMSE of 0.12 mV and an SNR of 4.68 dB.
\section{Material and Methods}

\subsection{Dataset Preprocessing}

In our experiments, we used the PTB-XL dataset \cite{ptbxl, physionet}, a large dataset of 21799 clinical 12-lead 10-second multi-annotated ECG signals from 18869 different patients.  Each raw ECG signal is digitally stored using 16 bit precision at a resolution of 1 $\mu$V per least significant bit (LSB) and with two available sampling frequencies: 100 Hz for low-resolution and 500 Hz for high-resolution. There also available 71 different SCP-ECG \cite{scpecg} conformed annotation statements covering both diagnostic and rhythm statements. We considered only the  annotations able to split the whole dataset into five diagnostic classes: Myocardial Infraction (MI), Conduction Disturbance (CD), Normal (NORM), Hypertrophy (HYP), ST/T Change (STTC). 

We preprocessed the dataset with the following filters:

\begin{itemize}
    \item an high-pass filter to remove components with frequency lower than 0.05 Hz from the low resolution 50 Hz signal;
    \item a band-pass filter to remove components outside the frequency range of 0.05-150 Hz from the high resolution 500 Hz signal \cite{aha1990standard, fuentes2012bpfilters}.
\end{itemize}

\begin{figure}[H]
    \centering
    \includegraphics[height=8cm,width=16cm]{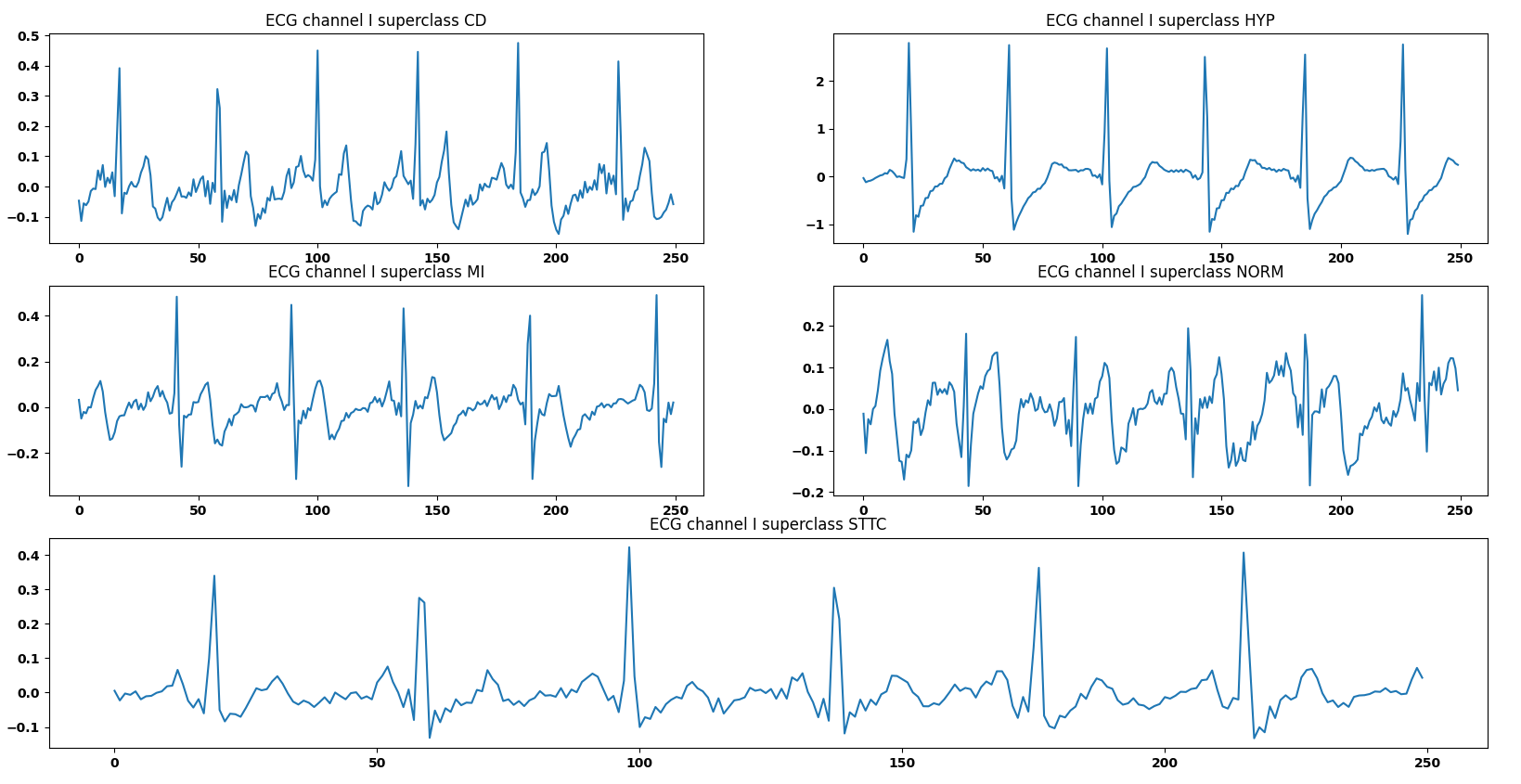}
    \caption{An example of ECG lead I channel LR data for each diagnostic superclass available in the PTB-XL dataset: Conduction Disturbance (CD), Hypertrophy (HYP), Myocardial Infraction (MI), Normal (NORM) and ST-T change (STTC)}
    \label{fig:figSignalsExample}
\end{figure}

We splitted each 10-second length filtered signal in two ECG windows with a fixed length of 5 seconds. Then, we downsampled each 100 Hz low-resolution filtered window to a sampling frequency of 50 Hz.  In Figure \ref{fig:figSignalsExample} we show an example of the lead I channel LR data for each of diagnostic superclass available in the dataset. 

We trained iteratively our autoencoder using a single class, and we used the remaining to evaluate its performances, and we splitted the dataset using 90\% of MI-related signals for training and the remaining 10\% for validation.

\begin{figure}[H]
    \centering
    \includegraphics[height=8cm,width=16cm]{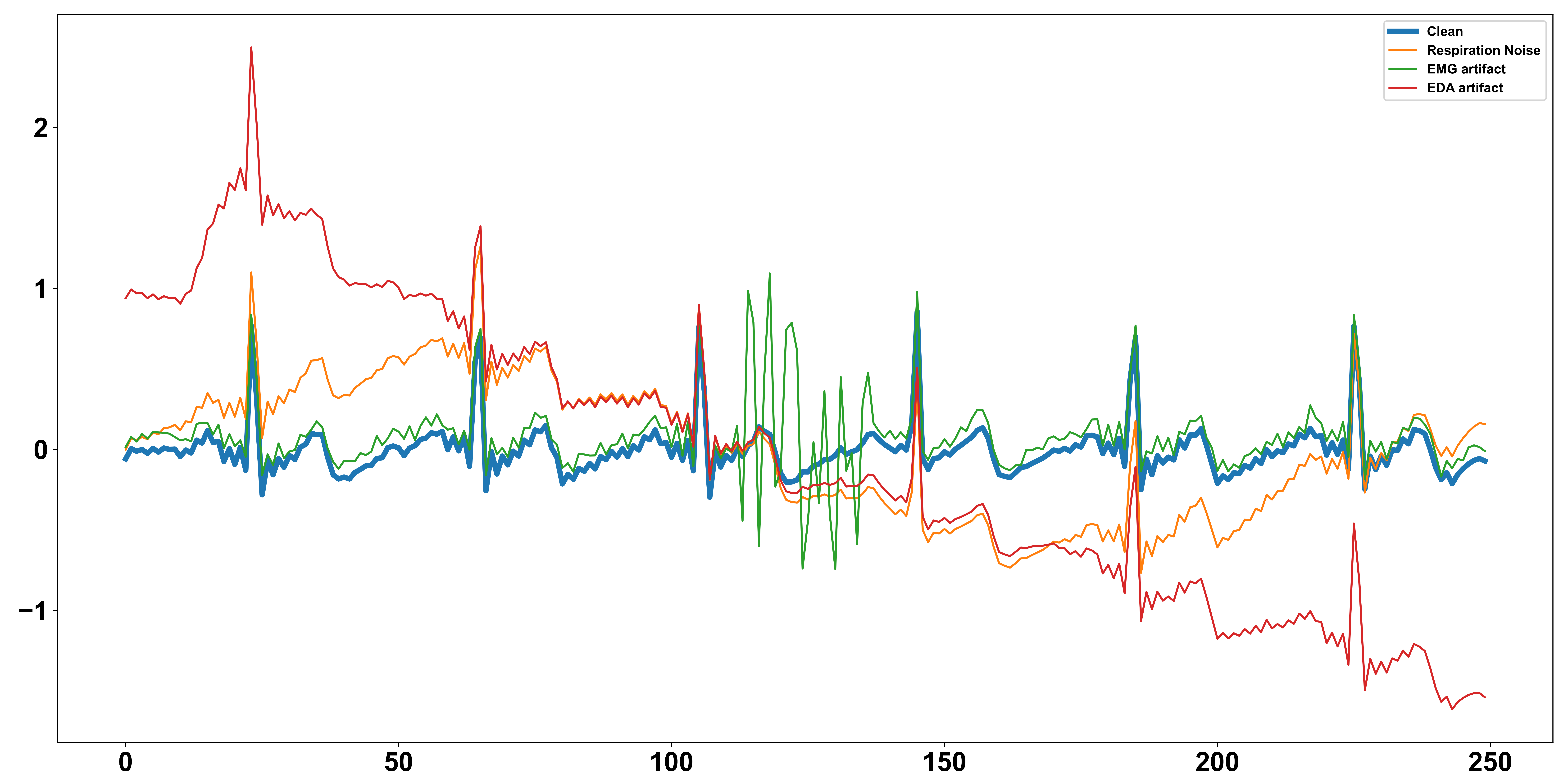}
    \caption{Example of not-corrupted lead I Myocardial Infraction LR ECG window and the three versions corrupted by respiration BW, EMG and EDA artifact respectively.}
    \label{fig:figartifact}
\end{figure}

Finally, we corrupted the LR training and test sets by adding synthetic ECG artifacts generated using the Neurokit2 Python library \cite{neurokit2}. A total of three different types of common ECG artifacts where generated: EMG, EDA, and BW artifact. For each signal, we used a randomized approach to choose whether the corruption should be applied or not, and in case of positive response, a second randomized approach was used to choose the artifact to add.  An example of a not-corrupted lead I Myocardial Infraction ECG LR window and the three versions of the same signal corrupted by different artifact are reported in Figure \ref{fig:figartifact}.

\subsection{The architecture of the DCAE-SR}

\begin{figure}[H]
    \centering
    \includegraphics[height=9cm,width=16cm]{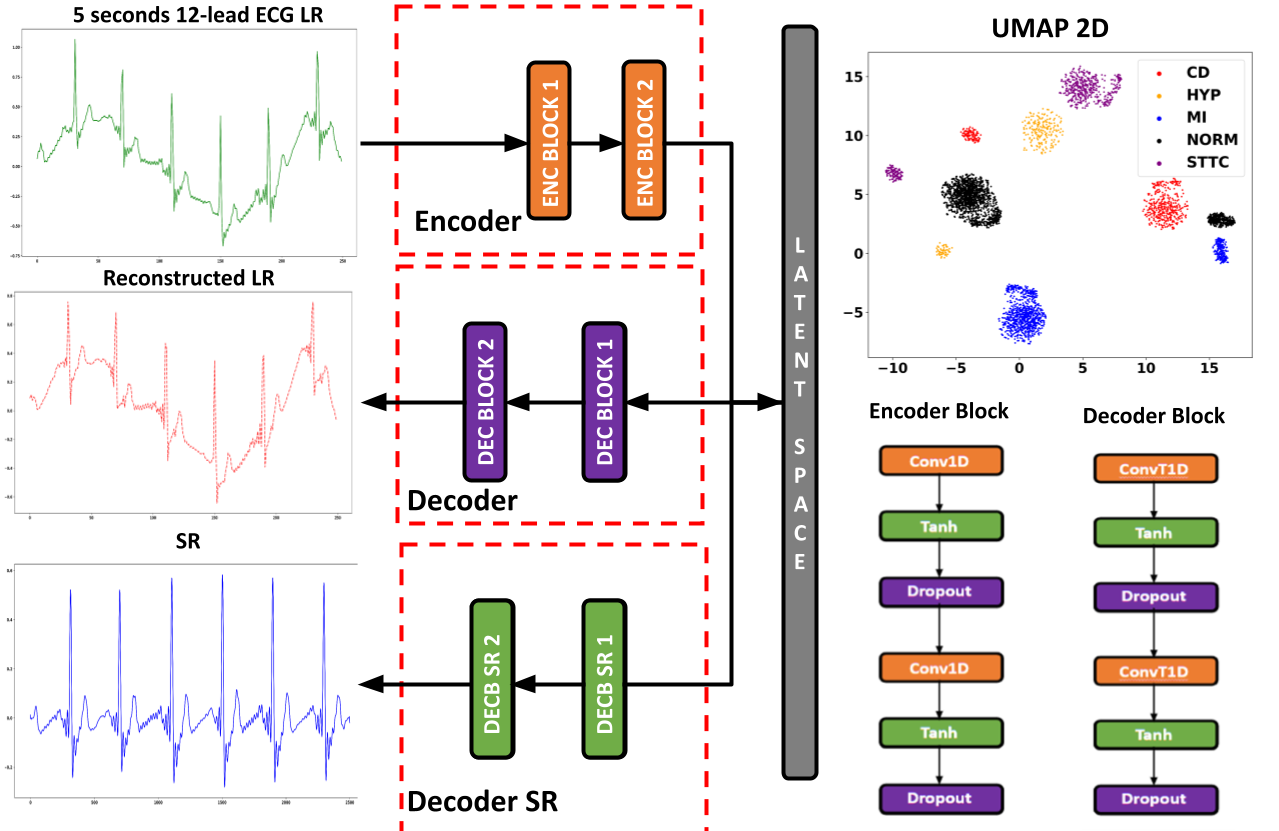}
    \caption{Our proposed proposed DCAE-SR architecture, with details about Convolutional blocks used to construct the Encoder and the two Decoders. We also report the 2D projection of the latent space obtained using the T-distributed Stochastic Neighbour Embedding (TSNE) on the latent space representation of the corrupted test set.}
    \label{fig:figArchitecture}
\end{figure}

Here, we propose our temporal-based denoising CAE for ECG high-resolution reconstruction (or ECG super-resolution), whose detailed architecture and the 2-dimensional projection of the latent space are depicted in Figure \ref{fig:figArchitecture}.  The hyperparameters obtained for the Encoder, the Decoder for reconstruction, and the Decoder for super-resolution (Decoder SR) are summarized in Table \ref{tableHyperparams}.

\begin{table}[H]

\small
\centering
\begin{tabular}{|c|c|c|c|c|c|} 
\hline
\textbf{Block} &\textbf{In channels} &\textbf{Out channels}& \textbf{Kernel size} & \textbf{Stride} &\textbf{Dropout}\\
\hline			
Encoder Block 1 & [12, 12] & [12, 192] & [3, 3] & [1, 1] & [0.1, 0.1]\\
\hline
Encoder Block 2 & [192, 384] & [384, 768] & [3, 3] & [1, 1] & [0.1, 0.1]\\
\hline
Decoder Block 1 & [768, 768] & [768, 384] & [3, 3] & [1, 1] & [0.1, 0.1]\\
\hline
Decoder Block 2 & [384, 192] & [192, 12] & [3, 3] & [1, 1] & [0.1, None]\\
\hline
DecoderBlock SR1 & [768, 768] & [768, 384] & [30, 30] & [5, 2] & [0.1, 0.1]\\
\hline
DecoderBlock SR2 & [384, 192] & [192, 12] & [10, 4] & [1, 1] & [0.1, None]\\
\hline
\end{tabular}
\caption{\label{tableHyperparams} Hyperparameters used for our DCAE-SR model.}
\end{table}

During the training phase, the model will receive a list containing the following input data:
\begin{itemize}
    \item 5-second length 50 Hz Low Resolution (LR) ECG window that goes inside the Encoder and mapped into the latent space;
    \item 5 second length 500 Hz High Resolution (HR) ECG window, that is the high resolution target. We need to specify that the HR input does not go inside the Encoder but is only used to perform the super-resolution error (also called high-resolution reconstruction loss) optimization.
\end{itemize}

The model tries to reconstruct both LR and HR signals using only the information retrieved in the LR one and mapped into the latent space. During the training, a multiple-loss function optimization was applied to optimize both the reconstruction, denoising, and super-resolution of the signal. To do that, the following mean squared errors (MSEs) were computed:

\begin{itemize}
    \item Reconstruction error: MSE between the LR signal and LR reconstruction in output of the Decoder;
    \item Super-resolution error: MSE between the target HR signal and the super-resoluted signal obtained in output of the Decoder-SR.
\end{itemize}

\begin{figure}[H]
    \centering
    
    \begin{subfigure}{1.0\textwidth}
    \includegraphics[height=5cm,width=18cm]{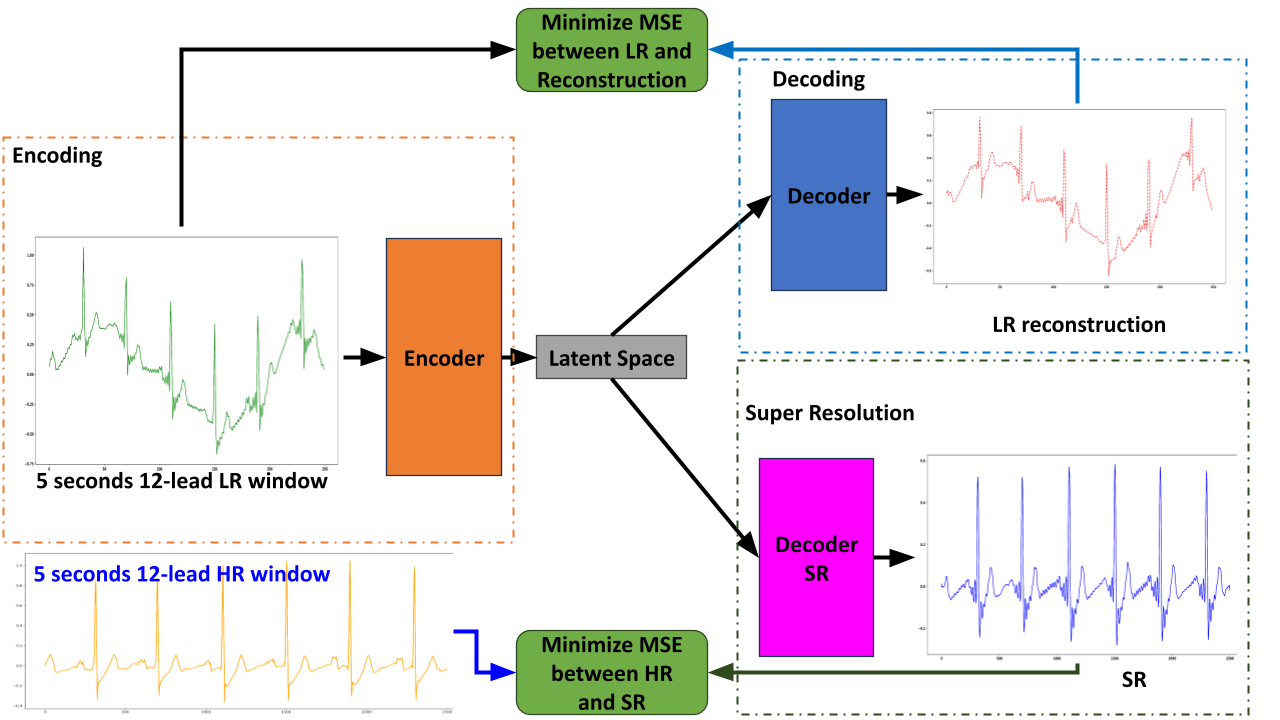}
    \caption{}
    \end{subfigure}

    \begin{subfigure}{1.0\textwidth}
    \includegraphics[height=4.5cm,width=18cm] {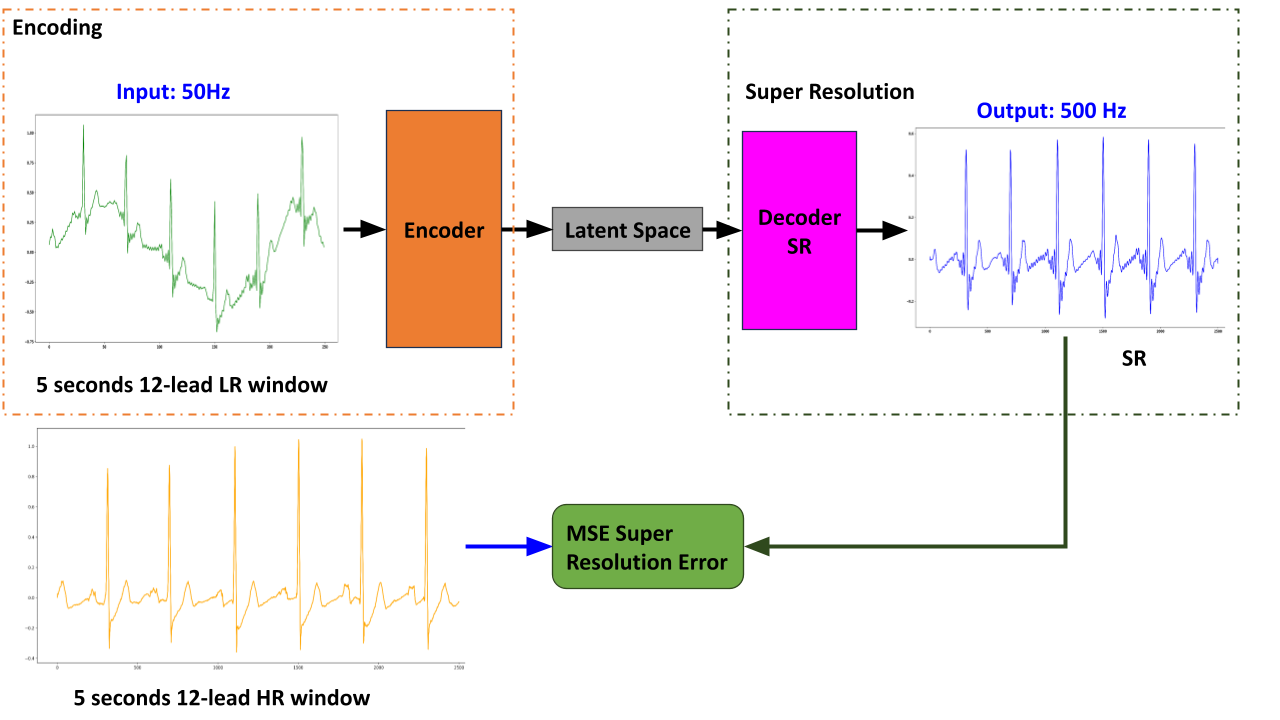}
    \caption{}
    \end{subfigure}
    
    \caption{Workflow of our super-resolution model during training phase (a) and validation phase (b)}
    \label{fig:figWorkflows}
\end{figure}

Training and validation phase workflows are depicted in Figure \ref{fig:figWorkflows}. The model was trained using a multi-loss optimization with one Adam optimizer, 0.0001 learning rate and 1 batch size for a total of 20 epochs.  

Multiple ablation studies were performed to identify the best combination of components regarding multiple/single losses optimization, with/without (w/wo) artifact presence in the LR signal, w/wo denoising power, w/wo super-resolution decoder, w/wo last Tanh activation function inside the last block of each decoder.

\begin{figure}[H]
    \centering
    \includegraphics[height=8cm, width=15cm]{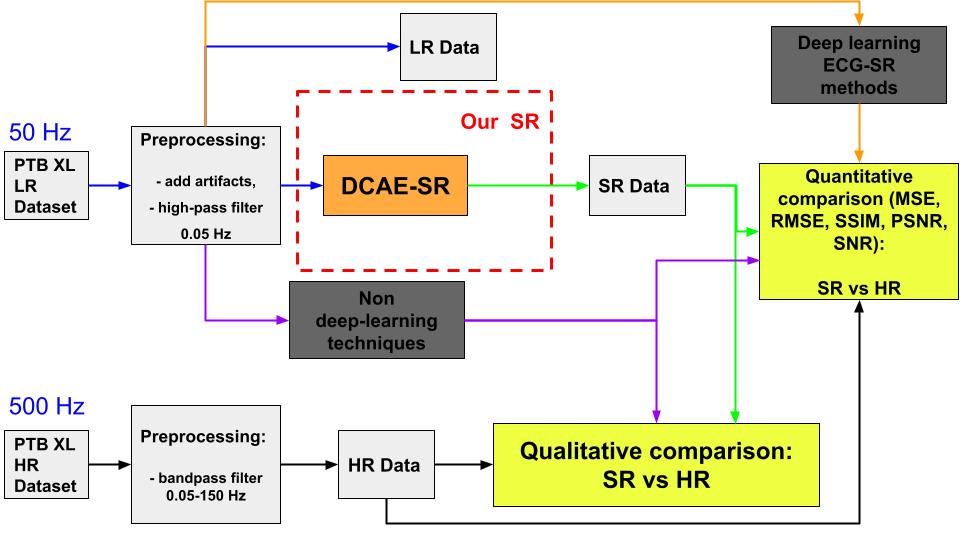}
    \caption{Workflow for qualitative and quantitative comparison of our super-resolution results against non-deep learning and deep learning based ECG super-resolution models available in literature.}
    \label{fig:figGeneralWf}
\end{figure}

We compared our super-resolution results against the ECG super-resolution deep learning methods available in the literature and other non-deep learning methods such as ECG bandpass filtering (for denoising) followed by cubic interpolation algorithm (for super-resolution). Results were compared using both qualitative and quantitative approaches. The general workflow used to compare our results is reported in Figure \ref{fig:figGeneralWf}.

\subsection{Metrics}

We assessed the performances of our DCAE-SR in  in both super-resolution and denoising tasks with respect to state-of-the-art methods by using the following quantitative metrics:

\begin{itemize}
    
    \item Mean Squared Error (MSE) \cite{mse} between the predicted 500 Hz super-resoluted signal $x_i$ and the ground truth 500 Hz HR one $y_i$. MSE measures the average squared sample-wise   difference between the ground truth and the predicted SR signals;
    \begin{equation}
    {
    \text{MSE($x_i$, $y_i$)} = \frac{1}{N} \sum_{i=1}^{N} (x_i - y_i)^2;
    }
    \end{equation}
    \textit{Where N is the number of ECG signals.}
    
    \item Root Mean Squared Error (RMSE) \cite{rmse}, between the predicted super-resoluted signal $x_i$ and the HR ground truth signal $y_i$, simply defined as the square root of the MSE.
    \begin{equation}
    {
    \text{RMSE($x_i$, $y_i$)} = \sqrt{MSE(x_i, y_i)}
    }
    \end{equation}
    
    \item Structural Similarity Index Measure (SSIM) \cite{ssim} is a metric that quantifies the structural similarity between the HR ground truth signal $y$ and the predicted super-resoluted signal $x$;
    \begin{equation}
    \text{SSIM(x, y)} = \frac{{(2 \mu_x \mu_y + C_1)(2 \sigma_{xy} + C_2)}}{{(\mu_x^2 + \mu_y^2 + C_1)(\sigma_x^2 + \sigma_y^2 + C_2)}}
    \end{equation}
    \textit{Where: $\mu_x$ and $\sigma_x$ are the mean and standard deviation values over a window in the super-resoluted signal $x$,  $\mu_y$ and $\sigma_y$ are the mean and standard deviation values over a window in the high-resolution signal $y$, $\sigma_{xy}$ is the covariance over a window between the super-resoluted signal $x$ and the high-resolution reference $y$. C1, C2, and C3 are three constant values.}

    \item Signal-to-Noise ratio (SNR) \cite{snr} between the predicted super-resoluted signal $x_i$ and the ground truth HR one $y_i$;
    \begin{equation}{
    \text{SNR($x_i$, $y_i$)} = 10 \cdot \log_{10}\left(\frac{\sum_{i=1}^{N} x_i^2}{\sum_{i=1}^{N} (x_i - y_i)^2}\right)
    }
    \end{equation}
    \textit{Where N is the number of ECG signals.}
    
    \item Peak Signal-to-Noise Ratio (PSNR) \cite{psnr} between the predicted super-resoluted signal $x_i$ and the ground truth HR signal $y_i$. PSNR measures the ratio between the maximum possible power of a signal and the power of corrupting noise. In the field of super resolution, it can be used to assess the quality of the super resolution;
    \begin{equation}{
    \text{PSNR($x_i$, $y_i$)} = 10 \cdot \log_{10} \frac{max(x_i)^2}{MSE(x_i, y_i)}
    }
    \end{equation}
    
\end{itemize}

\section{Results and Discussion}

\begin{figure}[H]
    \centering
    \includegraphics[height=10cm, width=16cm]{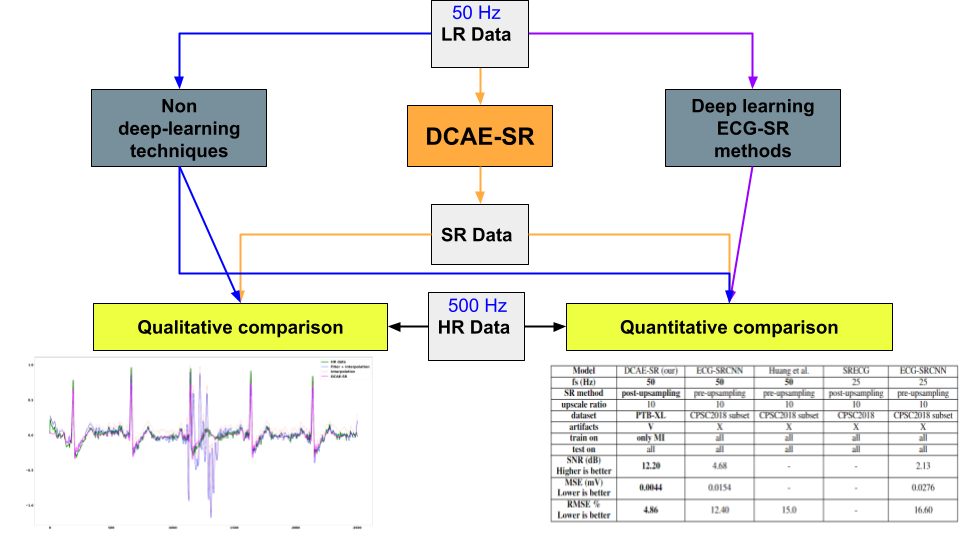}
    \caption{Workflow used to validate and compare our model DCAE-SR against deep and non-deep learning techniques for ECG high-resolution reconstruction from low-resolution corrupted signals. Examples of qualitative and quantitative comparisons are depicted.}
    \label{fig:figResults}
\end{figure}

This section presents the results of the use for reconstructing a HR from an input LR signal on the PTB-XL dataset. We compared these results with other state-of-the-art ECG super-resolution models and with a baseline of non-deep methods for obtaining super-resolution based on: (i) cubic interpolation; (ii) bandpass filter + cubic interpolation. In particular we measured both the quality of reconstruction of the HR signals and the ability to remove artifacts.
Figure \ref{fig:figResults} depicts the implement workflow of tests.

\begin{figure}[H]
    \centering
    \includegraphics[height=12cm, width=18cm]{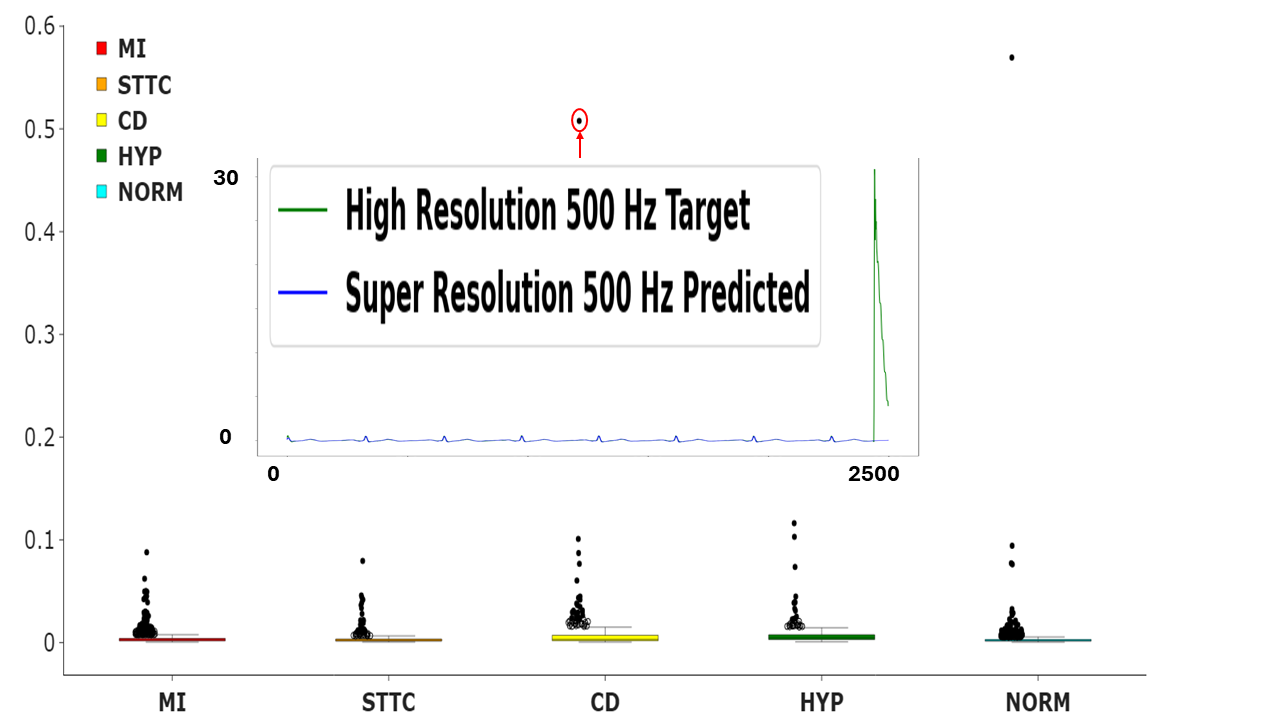}
    \caption{The right part of the figure shows the boxplot diagrams of super-resolution MSE error made by our DCAE-SR using all five diagnostic superclasses available in the corrupted test set. The left part of the Figure visualises the corresponding signals, showing the cause of the high MSE.}
    \label{fig:figBoxOut}
\end{figure}

First we measured the MSE super-resolution error obtained by our DCAE-SR in the super-resolution considering all the five different diagnostic superclasses as summarised in Table \ref{tableDCAEresults}. The relatively high values of maximum MSE for some classes may be explained by the presence of some outliers peaks in the signal. Figure \ref{fig:figBoxOut} shows boxplot containing the MSE super-resolution error distribution for each diagnostic superclass (on the right) and a zoomed visualization of high-resolution (in green) and super-resolution (in blue) signals with one of the highest MSE super-resolution error value (on the left).  The boxplot diagrams evidence a few outliers that negatively affect the mean value of MSE errors. Moreover, the visualisation of the signal corresponding to the outlier, as evidenced on the left part of the figure, shows the presence of an exceptionally high peak at the end of the V6 lead high-resolution target signal.  This confirms that some problems in our SR prediction do not cause the error in MSE. Actually, in those cases, the uncorrupted SR obtained using our approach can be considered of better quality.

\begin{table}
\large
\centering
\begin{tabular}{|c|c|c|c|} 
\hline
\textbf{Superclass} & \textbf{Min MSE} & \textbf{Max MSE} & \textbf{ Mean MSE (+/- STD)}\\
\hline			
CD	& 0.00044 & 0.5077 & 0.0081 (+/-) 0.0266\\
\hline
HYP	& 0.0007 & 0.1162 & 0.0079 (+/-) 0.0135\\
\hline
NORM & 0.0004 & 0.5694 & 0.0030 (+/-) 0.0058\\
\hline
MI & 0.0004 & 0.0879 & 0.0040 (+/-) 0.0060\\
\hline
STTC & 0.0005 & 0.1162 & 0.0034 (+/-) 0.0058 \\
\hline
\end{tabular}
\caption{\label{tableDCAEresults} For each class of the input signal, we show the results in terms of minimum, maximum, and mean $\pm$ standard deviation. }
\end{table}

\begin{figure}[H]
    \centering
    \includegraphics[height=10cm, width=17cm]{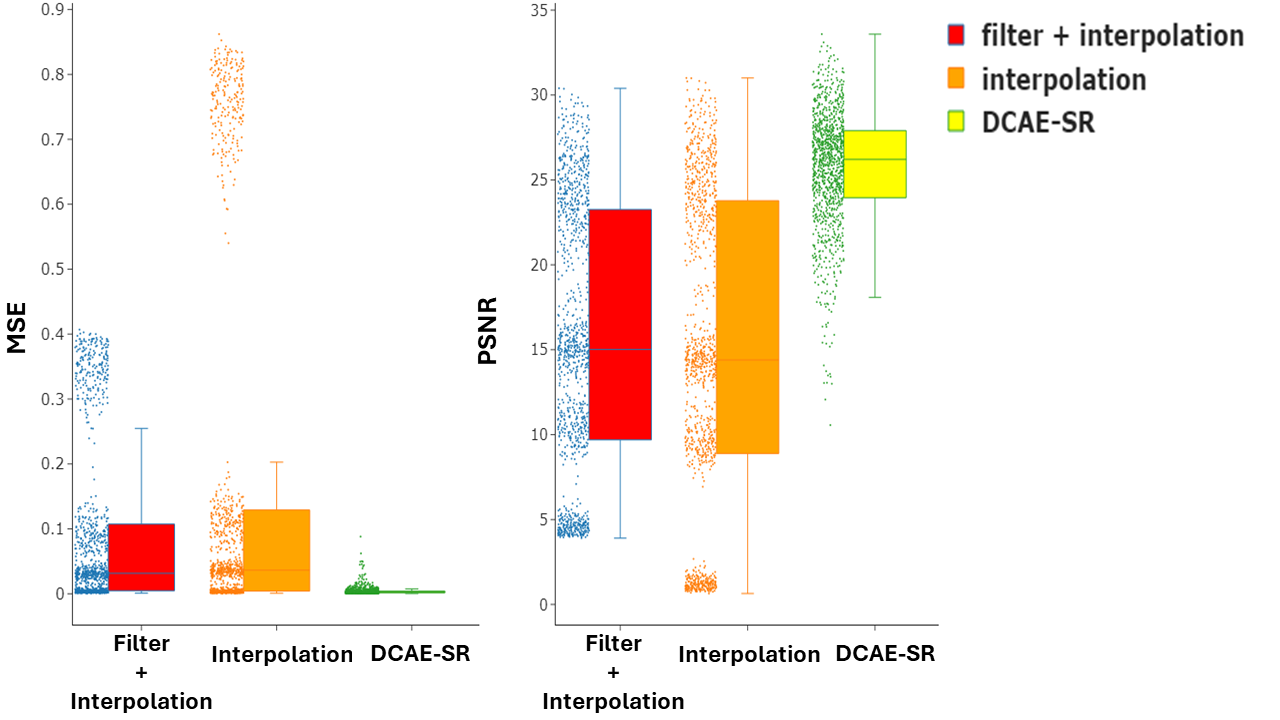}
    \caption{Boxplot of MSE values (left) and PSNR values (right) obtained with our proposed DCAE-SR method and other non deep learning algorithms using the corrupted test set of the MI superclass.}
    \label{fig:figBoxComp}
\end{figure}

We now compare the MSE and the PSNR of our model with baseline non-deep learning super-resolution methods by considering only the Myocardial Infarction (MI) class as depicted in Figure \ref{fig:figBoxComp}. 

\begin{figure}[H]
    \centering
    \includegraphics[height=8cm, width=16cm]{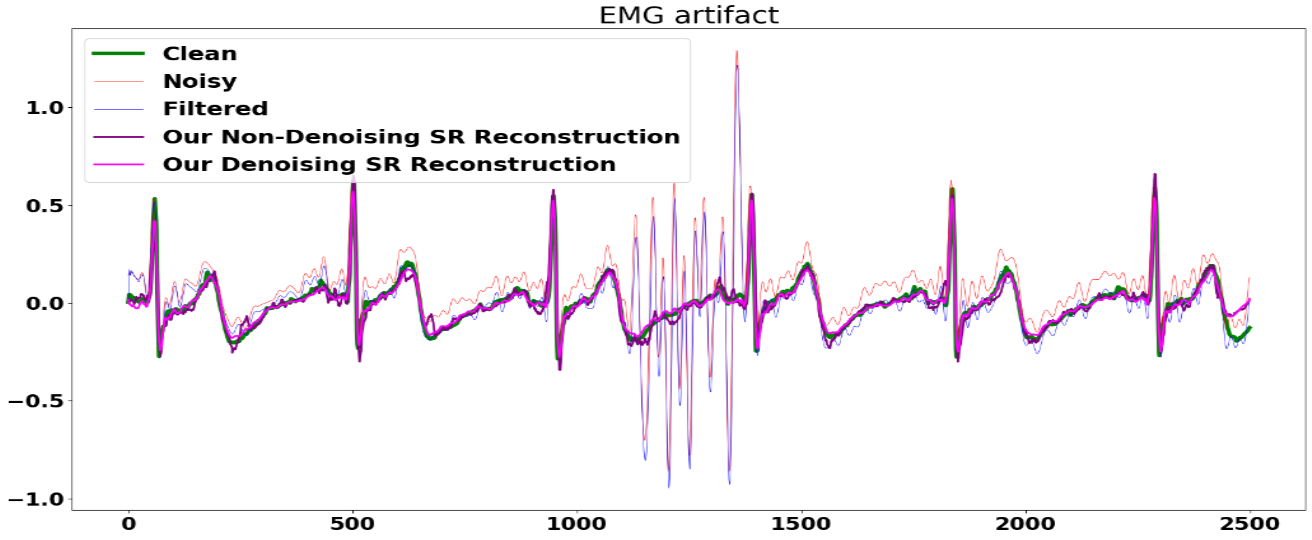}
    \caption{Qualitative comparison of our denoising and non-denoising super-resolution model against the band-pass filter + cubic interpolation baseline.}
    \label{fig:figSREMG}
\end{figure}

Moreover, we also measured the performances of our model in reconstructing signals corrupted by high-frequency noise, which represents a major challenge. Figure \ref{fig:figSREMG}  reports a qualitative comparison of our super-resolution methods when it receives an LR myocardial infraction ECG signal corrupted by EMG artifact as input. The EMG signal bandwidth (20-500 Hz) partially overlaps with the ECG bandwidth (0.05-150 Hz), so, for example, a classic bandpass filter does not work well in this case, as we can see in the figure mentioned. Our denoising model, on the other hand, can distinguish noise/artifact patterns inside the latent space and remove them during 
super-resolution.

\begin{figure}[H]
    \centering
    \includegraphics[height=6cm, width=16cm]{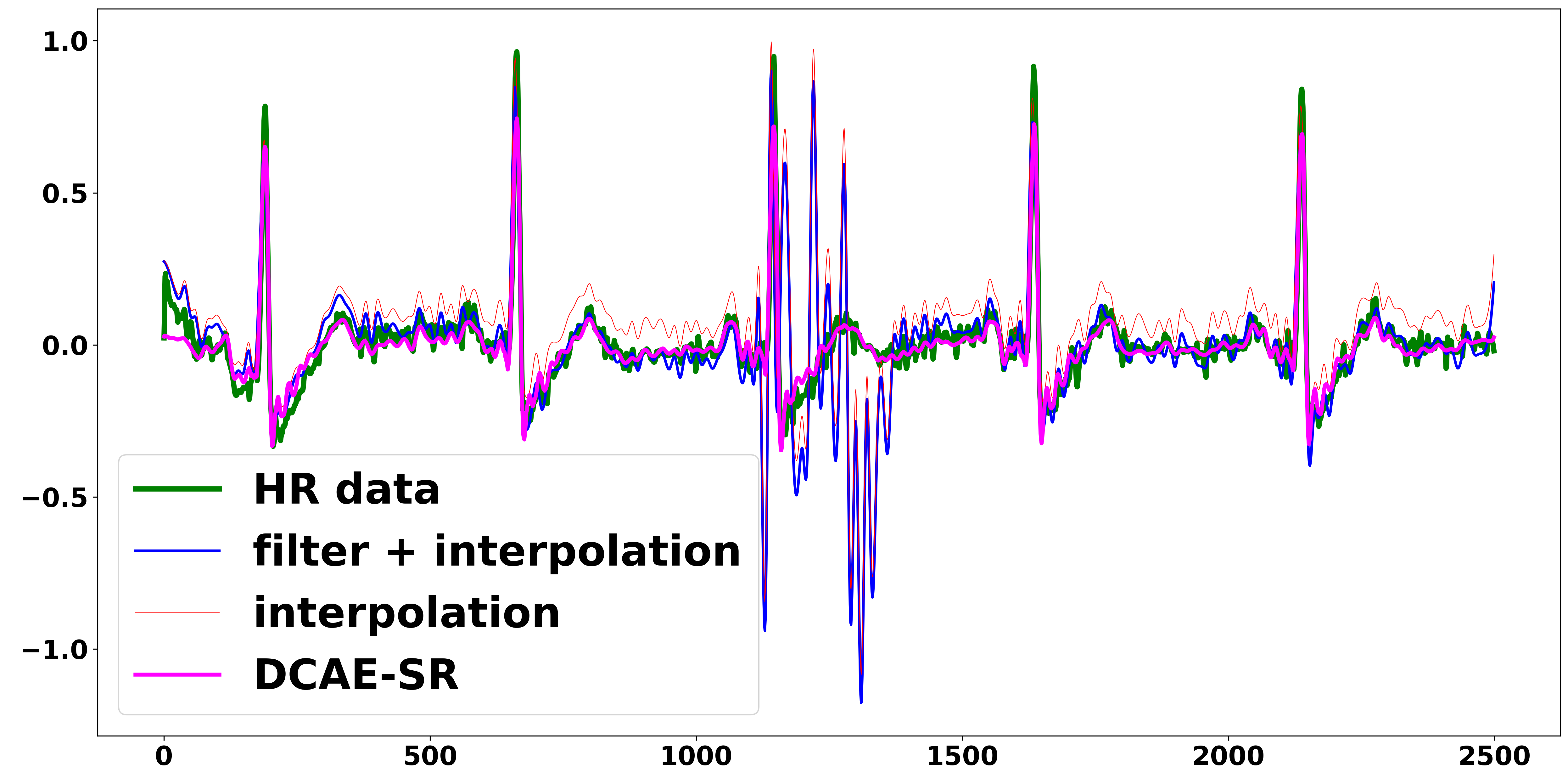}
    \caption{Qualitative comparison of our DCAE-SR model with other reproducible non deep learning denoising super-resolution methods using a myocardial infraction ECG signal corrupted by EMG artifact.}
    \label{fig:figSREMGcomp}
\end{figure}

Using the same corrupted signal, in Figure \ref{fig:figSREMG} we report the difference in qualitative results obtained using our models with/without denoising power. The model without denoising power struggles (but not to much) in the super-resolution due to the presence of the EMG artifact. 

We also tested the performance of dealing with missing channels inside the signal. Missing channels were simulated using a random approach, and we computed the performances regarding MSE super-resolution errors obtained on the corrupted MI superclass test set. For each channel and each signal in the test set, a random number between 0 and 1 is sampled; if the number is greater or equal to a certain threshold $p$ (called missing channel rate), then the channel is considered missing and replaced by an all zeros tensor. For this experiment, a maximum of one channel per signal can be missing.

In Table \ref{tableDCAEChannelDeletion0.2} we report changes regarding minimum, maximum, and mean MSE super-resolution values of a single channel missing using a missing rate $p$ of 0.2. The table is ordered in descending order in respect of mean MSE value. As we can see, the aVL channel seems to be the most important lead for the super-resolution of a MI signal: this is proven by the lowest super-resolution performance obtained when this channel is removed.

\begin{table}[H]
\centering
\normalsize
\begin{tabular}{|c|c|c|c|} 
\hline
\textbf{Missing Channel} & \textbf{Min MSE} & \textbf{Max MSE} & \textbf{	Mean MSE (+/- STD)}\\
\hline
aVL	& 0.00049	& 0.1476	& 0.00793(+/-)0.01555\\
\hline
I	& 0.00044	& 0.13584	& 0.00752(+/-)0.01407\\
\hline
V3	& 0.00044	& 0.12419	& 0.00732(+/-)0.01304\\
\hline
aVR	& 0.00044	& 0.08788	& 0.00629(+/-)0.01062\\
\hline
V2	& 0.00044	& 0.15302	& 0.00624(+/-)0.01023\\
\hline
V5	& 0.00044	& 0.08788	& 0.00616(+/-)0.00915\\
\hline
III	& 0.00044	& 0.11793	& 0.00605(+/-)0.01011\\
\hline
aVF	& 0.00044	& 0.08788	& 0.00563(+/-)0.00827\\
\hline
V4	& 0.00044	& 0.08788	& 0.00541(+/-)0.00801\\
\hline
V1	& 0.00044	& 0.08788	& 0.00513(+/-)0.00721\\
\hline
II	& 0.00049	& 0.10501	& 0.00509(+/-)0.00719\\
\hline
V6	& 0.00049	& 0.08788	& 0.0046(+/-)0.00638\\
\hline
No Channel Missing & 0.0004 & 0.0879 &	0.0040 (+/-)0.0060\\
\hline
\end{tabular}
\caption{\label{tableDCAEChannelDeletion0.2} Super resolution performances of the DCAE-SR model in case of missing single channel during testing on noised MI ECG signals. The results of this  table are obtained using a missing channel rate of 0.2 (20\%). The table is ordered by "Mean MSE" values, higher values mean worst super resolution if the channel is missing.} 
\end{table}

We then tried to change the aVL missing rate $p$ in the range of 0 to 1 and report the results of this experiment in Table \ref{tableaVFmissingPs}.

\begin{table}[H]

\centering
\large
\begin{tabular}{|c|c|c|c|} 
\hline
\textbf{Missing Rate of Channel aVL} & \textbf{Min MSE} & \textbf{Max MSE} & \textbf{Mean MSE (+/- STD)}\\
\hline
0 & 0.0004 & 0.0879 &	0.0040 (+/-)0.0060\\
\hline
0.1	& 0.00044 & 0.08788	& 0.00629(+/-)0.01254\\
\hline
0.2	& 0.00049 & 0.1476 & 0.00793(+/-)0.01555\\
\hline
0.3	& 0.00044 & 0.10527	& 0.0101(+/-)0.01854\\
\hline
0.4	& 0.00044 & 0.09236	& 0.01209(+/-)0.02048\\
\hline
0.5	& 0.00044 & 0.1476	& 0.01286(+/-)0.02124\\
\hline
0.6	& 0.00049 & 0.09786	& 0.01531(+/-)0.02313\\
\hline
0.7	& 0.00049 & 0.1476	& 0.01711(+/-)0.02463\\
\hline
0.8	& 0.00049 & 0.1476	& 0.02019(+/-)0.02655\\
\hline
0.9	& 0.00044 & 0.1476	& 0.02112(+/-)0.02656\\
\hline
\end{tabular}
\caption{\label{tableaVFmissingPs} Super-resolution performances of our DCAE-SR model in the case of different values of the aVL missing rate $p$.}
\end{table}

We quantitatively compared our super-resolution performances against other deep learning related works as reported in Table \ref{tableRelated}. Only results regarding an upscale ratio of 10 and a low-resolution signal with sampling rate equal or lower than 50 Hz are reported. Of course, using a higher low-resolution sampling rate (for example 100 Hz) corresponds to an higher bandwidth and a lower upscale ratio (for example 5): fewer points to predict that will lead to better performance in terms of super-resolution.

Here we summarize the main differences between us and related works:
\begin{itemize}
    \item we used a different dataset, the PTB-XL, because it natively provide both low resolution (100 Hz) and high resolution (500 Hz) signals. It also provide a standard train-test split for a fair comparison between different deep-learning models;
    \item we are the first to take in consideration the presence of noise and artifacts inside the ECG signal, and their impact in the overall super-resolution performances;
    \item we used a different training approach for the ECG super-resolution task. The model is trained on only one particular diagnostic superclass (myocardial infarction) and then tested on all five superclasses to see how the model generalise on the unseen types of cardiac pathologies; 
    \item quantitative performances obtained depicts our model as the current state-of-the-art (SOTA) model for the super-resolution of very low-resolution ECG signals (50 Hz). SOTA in both PSNR, SNR, MSE, SSIM and RMSE super-resolution metrics, computed between the super-resoluted signal predicted by our model and the target high resolution signal available.
\end{itemize}

\begin{table}[H]
\centering
\normalsize
\begin{tabular}{|c|c|c|c|c|c|} 
\hline
\textbf{Model} & DCAE-SR (our) & ECG-SRCNN & Huang et al. & SRECG & ECG-SRCNN \\
\hline
\textbf{fs (Hz)} & \textbf{50} &  \textbf{50} &  \textbf{50} & 25 & 25 \\
\hline 
\textbf{SR method} & \textbf{post-upsampling} & pre-upsampling & pre-upsampling & post-upsampling & pre-upsampling\\
\hline
\textbf{upscale ratio} & 10 & 10 & 10 & 10 & 10 \\
\hline
\textbf{dataset} & \textbf{PTB-XL} & CPSC2018 subset  & CPSC2018 subset  & CPSC2018  & CPSC2018 subset \\
\hline
\textbf{artifacts} & \textbf{V} & X & X & X & X \\
\hline
\textbf{train on} & \textbf{only MI} & all  & all & all & all \\
\hline
\textbf{test on} & all & all & all & all & all \\
\hline 
\makecell{ \textbf{SNR (dB)} \\ \textbf{Higher is better}} & \textbf{12.20} & 4.68 & -  & -  & 2.13 \\
\hline
\makecell{ \textbf{MSE (mV)} \\ \textbf{Lower is better}} & \textbf{0.0044} & 0.0154  & -  & -  & 0.0276 \\
\hline
\makecell{ \textbf{RMSE \%} \\ \textbf{Lower is better}} & \textbf{4.86} & 12.40  & 15.0  & -  & 16.60 \\ 
\hline
\end{tabular}
\caption{\label{tableRelated} Comparison of our proposed method with other super-resolution ECG methods. We reported only super resolution performances on very low resolution signals (sampling rate lower or equal then 50 Hz).}
\end{table}

\begin{figure}[H]
    \centering
    \includegraphics[height = 8cm, width = 16cm]{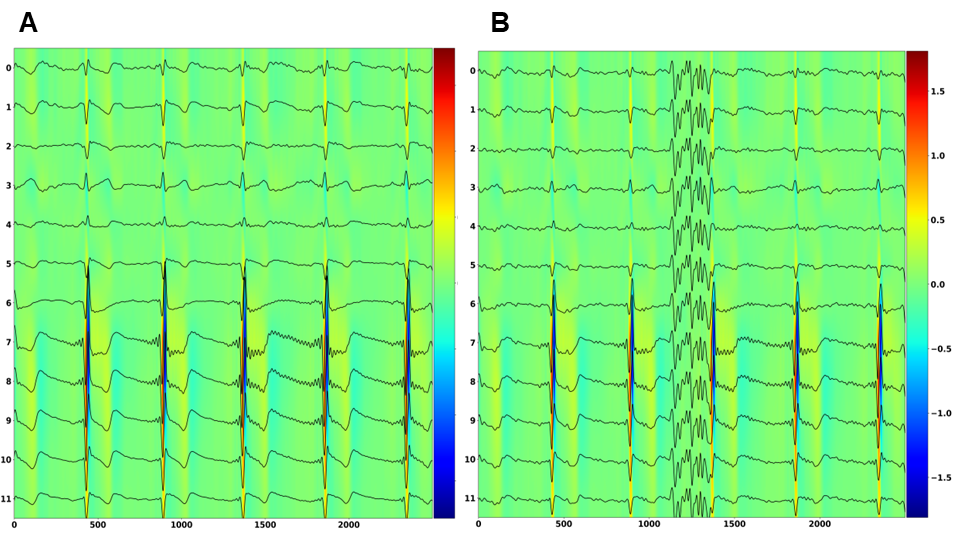}
    \caption{Activation maps obtained by our DCAE-SR for the super-resolution of the same ECG window without any corruption (A) and corrupted by EMG artifact (B).}
    \label{fig:figActivationmapsNoise}
\end{figure}

\begin{figure}[H]
    \centering
    \includegraphics[height = 8cm, width = 16cm]{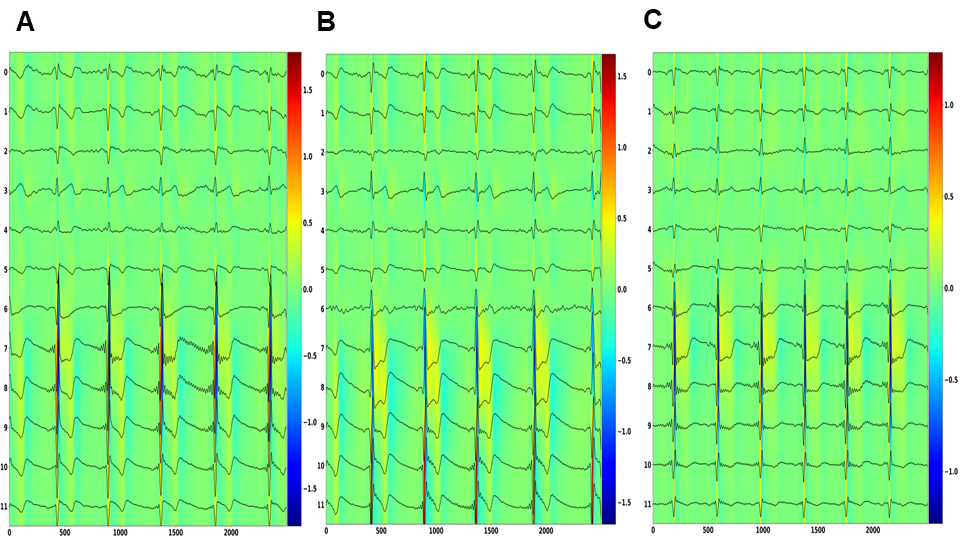}
    \caption{Activation maps obtained by our DCAE-SR for the super-resolution of the three uncorrupted ECG windows with different diagnostic superclass: (A) Normal, (B) Myocardial Infraction, (C) ST-T change}
    \label{fig:figActivationmaps}
\end{figure}

Finally, we report some explainability of the super-resolution performed by our DCAE-SR model. Figure \ref{fig:figActivationmapsNoise} report the activation maps obtained by using our model to super-resolute a Normal (NORM) ECG window corrupted by EMG artifact (B) and without corruption (A). Together with the activation map, we report all twelve leads of the interpolated low-resolution ECG signal. As we can see, the two activation maps are very similar, proving the robustness of our model in case of corruption inside the low-resolution ECG. At the same time, our model is also capable of correctly detect the R peak hidden inside the EMG artifact and delineating the entire PQRST beat, opening a new case of use of our model: the task of R-peak detection and PQRST delineation. Figure \ref{fig:figActivationmaps} shows the activation maps obtained during the super-resolution of three ECG windows with different diagnostic superclasses: normal (A), myocardial infraction (B) and ST-T change (C), respectively. 

\subsection{Ablation studies}

We tested our model and trained on corrupted ECG signals to perform super-resolution in the clean version of the test set (version without the synthetically generated artifacts). This was done to prove our model's efficiency in both corrupted and unnoised conditions. This was only one of the ablation studies performed to investigate the specific contributions/effects of individual components within the experimental setup. 

Table \ref{tableSRDenoising} shows performances on the corrupted test set of our DCAE-SR model with/without Denoising power or w/wo super-resolution Decoder. In the case of a no super-resolution Decoder, the temporal information is augmented on the low-resolution reconstruction using a cubic interpolation layer with an upsampling rate of 10. Using denoising and super-resolution together gives better performance in the super-resolution, both quantitatively and qualitatively.

Table \ref{tableMultipleLosses} shows the ablation studies performed to choose the best combination of MSE losses optimization for the super-resolution task. The three possible combinations are low-resolution reconstruction (LR), high-resolution reconstruction (HR), and both low and high-resolution reconstruction (LR+HR). The double loss optimization gives the model a more explicit representation of the low-resolution input data into the latent space (thanks to the LR reconstruction optimization) and, simultaneously, a denoised and super-resolution representation of the same signal thanks to the HR reconstruction optimization.

Table \ref{tableCAEDCAE} shows performances of our super-resolution w/wo Denoising in ECG signals w/wo artifacts presence. The model can denoise the signal and shows similar performances in both corrupted and non-corrupted conditions, proving its strength in dealing with both environments.

Table \ref{tableDecBlockAblation} reports ablation studies performed to find the best architecture for the last block of the Decoder to obtain a better signal reconstruction and super-resolution.

\begin{table}[H]

\centering
\small
\begin{tabular}{|c|c|c|c|c|c|} 
\hline
\textbf{Denoising} & \textbf{Super-Resolution} &
\textbf{Mean MSE (+/- STD)} & 
\textbf{SSIM} &
\textbf{PSNR} & 
\textbf{Description}
\\
\hline			
X &	X &	0.1363 & 0.9169	& 14.1703 &  Reconstruction \\
\hline
V &	X &	0.0073 & 0.9939	& 22.9968 & Reconstruction + Denoising \\
\hline
X &	V &	0.0686 & 0.9410	& 16.7092 & Reconstruction + Super-Resolution \\
\hline
\textbf{V} & \textbf{V} &	\textbf{0.0058} & \textbf{0.9942} & \textbf{24.6254} & \textbf{Reconstruction + Denoising + Super-Resolution}\\
\hline
\end{tabular}
\caption{\label{tableSRDenoising} Ablation study to find the best combination of denoising and our Decoder-based super-resolution in case of noisy signals. Ablation studies performances computed only on the MI test set.}
\end{table}

\begin{table}[H]

\centering
\normalsize
\begin{tabular}{|c|c|c|c|} 
\hline 
\textbf{Loss type} & \textbf{MSE (lower is better)} & \textbf{SSIM (higher is better)} & \textbf{PSNR (higher is better)} \\
\hline
LR (optimize reconstruction) & 0.0073 & 0.9939 & 23.00 \\
\hline
HR (optimize super-resolution) & 0.0069 & 0.9935 & 23.45 \\
\hline
\textbf{LR+HR} (optimize both) & \textbf{0.0058} & \textbf{0.9942} & \textbf{24.62} \\
\hline
\end{tabular}
\caption{\label{tableMultipleLosses} Ablation studies on the multiple-loss optimization technique. The super-resolution by optimizing only the LR reconstruction was performed as follows: denoised LR reconstruction using our DCAE-SR decoder + cubic interpolation (see Table \ref{tableSRDenoising} with Denoising and without our Decoder for super-resolution). Ablation studies performances computed only on the MI test set.
}
\end{table}

\begin{table}[H]

\centering
\large
\begin{tabular}{|c|c|c|c|c|} 
\hline
\textbf{Denoising} & \textbf{Noise} & \textbf{MSE (lower is better)} & \textbf{SSIM (higher is better)} & \textbf{PSNR (higher is better)} \\
\hline
\textbf{True} & \textbf{True}	& \textbf{0.0058} &	\textbf{0.9942} & \textbf{24.62} \\
\hline
False	& True	& 0.0686	& 0.9410	& 16.71 \\
\hline
\textbf{True} & \textbf{False} &	\textbf{0.0054} & \textbf{ 0.9946} & \textbf{25.21} \\
\hline
False & False	& 0.0056	& 0.9941	& 24.77\\
\hline
\end{tabular}
\caption{\label{tableCAEDCAE} Performance of DCAE and CAE based super-resolution models in different cases of noise presence. Ablation studies performances computed only on the MI test set.}
\end{table}

\begin{table}[H]

\centering
\large
\begin{tabular}{|c|c|c|c|} 
\hline
\textbf{Last Tanh} & \textbf{Mean MSE } &  \textbf{SSIM} & \textbf{PSNR} \\
\hline			
\textbf{X} & \textbf{0.0040} & \textbf{0.9979} & \textbf{25.7013} \\
\hline 
V & 0.0058 & 0.9942 & 24.6254 \\
\hline
\end{tabular}
\caption{\label{tableDecBlockAblation} Ablation study to find the most efficient architecture for the last block of each decoder. Ablation studies performances computed only on the MI test set.}
\end{table}

\section{Conclusion}

This research paper introduces the Denoised Convolutional AutoEncoder (DCAE) for Electrocardiogram (ECG) super-resolution. The study focuses on improving ECG signals' resolution for better disease diagnosis and monitoring, using the PTB-XL dataset. The DCAE-SR model performs both super-resolution and denoising of ECG signals, showcasing superior performance over existing ECG super-resolution methods. The study highlights the critical role of individual ECG channels in super-resolution performance. Additionally, the inclusion of denoising capabilities within the super-resolution process improves the quality of the ECG signal and opens new avenues for applications. The study applies model explainability to shed light on the operational dynamics of the DCAE-SR model. The results lay a solid foundation for future research, including exploring the application of the model to other biomedical signals, integrating real-time processing capabilities, and further enhancing the model's explainability.

\section*{Declarations}

\subsection*{Availability of data and materials}
Data are available at: \url{https://physionet.org/content/ptb-xl/1.0.3/}.
Pretrained models and code used in this work are available upon request.

\subsection*{Declaration of Competing Interest}
Authors declare that they have no competing interests.

\subsection*{Funding}
UL Ph.D. fellow is partially funded by Relatech S.p.A. 

\subsection*{Contributions}
UL was responsible of software implementation, validation, testing, and editing the manuscript. PHG, PV and PL supervised all the processes and edited the manuscript.
All authors read and approved the manuscript.

\subsection*{Acknowledgements}
PHG has been partially supported by the Next Generation EU - Italian NRRP, Mission 4, Component 2, Investment 1.5, call for the creation and strengthening of 'Innovation Ecosystems', building 'Territorial R\&D Leaders' (Directorial Decree n. 2021/3277) - project Tech4You - Technologies for climate change adaptation and quality of life improvement, n. ECS0000009. 


\begin{thebibliography}{10}

\bibitem{prince2006medical}
Prince~Jerry L and Links~Jonathan M.
\newblock {\em Medical imaging signals and systems}, volume~37.
\newblock Pearson Prentice Hall Upper Saddle River, 2006.

\bibitem{meyer2004pattern}
Anke Meyer-B{\"a}se.
\newblock {\em Pattern Recognition and Signal Analysis in Medical Imaging}.
\newblock Academic Press, 2004.

\bibitem{satija2018review}
Udit Satija, Barathram Ramkumar, and M~Sabarimalai Manikandan.
\newblock A review of signal processing techniques for electrocardiogram signal
  quality assessment.
\newblock {\em IEEE reviews in biomedical engineering}, 11:36--52, 2018.

\bibitem{wasimuddin2020stages}
Wasimuddin Muhammad, Elleithy Khaled, Abuzneid Abdel-Shakour, Faezipour Miad,
  and Abuzaghleh Omar.
\newblock Stages-based ecg signal analysis from traditional signal processing
  to machine learning approaches: A survey.
\newblock {\em IEEE Access}, 8:177782--177803, 2020.

\bibitem{siontis2021artificial}
Konstantinos~C Siontis, Peter~A Noseworthy, Zachi~I Attia, and Paul~A Friedman.
\newblock Artificial intelligence-enhanced electrocardiography in
  cardiovascular disease management.
\newblock {\em Nature Reviews Cardiology}, 18(7):465--478, 2021.

\bibitem{ajdaraga2017fs}
Era Ajdaraga and Marjan Gusev.
\newblock Analysis of sampling frequency and resolution in ecg signals.
\newblock In {\em 2017 25th Telecommunication Forum (TELFOR)}, pages 1--4,
  2017.

\bibitem{kligfield2007prevalence}
Paul Kligfield and Peter~M Okin.
\newblock Prevalence and clinical implications of improper filter settings in
  routine electrocardiography.
\newblock {\em The American journal of cardiology}, 99(5):711--713, 2007.

\bibitem{garcia2009technical}
Javier Garc{\'\i}a-Niebla, Pablo Llontop-Garc{\'\i}a, Juan~Ignacio
  Valle-Racero, Guillem Serra-Autonell, Velislav~N Batchvarov, and
  Antonio~Bay{\'e}s De~Luna.
\newblock Technical mistakes during the acquisition of the electrocardiogram.
\newblock {\em Annals of Noninvasive Electrocardiology}, 14(4):389--403, 2009.

\bibitem{pizzuti1985digitalsampling}
G.P. Pizzuti, S.~Cifaldi, and G.~Nolfe.
\newblock Digital sampling rate and ecg analysis.
\newblock {\em Journal of Biomedical Engineering}, 7(3):247--250, 1985.

\bibitem{bui2021comparison}
Ngoc-Thang Bui and Gyung-su Byun.
\newblock The comparison features of ecg signal with different sampling
  frequencies and filter methods for real-time measurement.
\newblock {\em Symmetry}, 13(8):1461, 2021.

\bibitem{shekatkar2017detecting}
Snehal~M Shekatkar, Yamini Kotriwar, KP~Harikrishnan, and G~Ambika.
\newblock Detecting abnormality in heart dynamics from multifractal analysis of
  ecg signals.
\newblock {\em Scientific reports}, 7(1):15127, 2017.

\bibitem{narayanaswamy2002HR}
Narayanaswamy S.
\newblock High resolution electrocardiography.
\newblock {\em Indian Pacing Electrophysiol J.}, 2002.

\bibitem{berbari1998latepot}
Edward~J. Berbari and Ralph Lazzara.
\newblock {An Introduction to High-Resolution ECG Recordings of Cardiac Late
  Potentials}.
\newblock {\em Archives of Internal Medicine}, 148(8):1859--1863, 08 1988.

\bibitem{huang2023ecgsr}
Huang Rui, Xue Xiaojun, Xiao Renjie, and Bu~Fan.
\newblock A novel method for ecg signal compression and reconstruction:
  Down-sampling operation and signal-referenced network.
\newblock {\em Electronics}, 12(8), 2023.

\bibitem{kaniraja2024deep}
Christina~Perinbam Kaniraja and Deepak Mishra.
\newblock A deep learning framework for electrocardiogram (ecg) super
  resolution and arrhythmia classification.
\newblock {\em Research on Biomedical Engineering}, pages 1--13, 2024.

\bibitem{chen2023srecg}
Tsai-Min Chen, Yuan-Hong Tsai, Huan-Hsin Tseng, Kai-Chun Liu, Jhih-Yu Chen,
  Chih-Han Huang, Guo-Yuan Li, Chun-Yen Shen, and Yu~Tsao.
\newblock Srecg: Ecg signal super-resolution framework for portable/wearable
  devices in cardiac arrhythmias classification.
\newblock {\em IEEE Transactions on Consumer Electronics}, 2023.

\bibitem{arsene2019deep}
Arsene~Corneliu TC, Hankins Richard, and Yin Hujun.
\newblock Deep learning models for denoising ecg signals.
\newblock In {\em 2019 27th European signal processing conference (EUSIPCO)},
  pages 1--5. IEEE, 2019.

\bibitem{wu2020extracting}
Wu~Xiaodan, Zheng Yumeng, Chu Chao-Hsien, and He~Zhen.
\newblock Extracting deep features from short ecg signals for early atrial
  fibrillation detection.
\newblock {\em Artificial Intelligence in Medicine}, 109:101896, 2020.

\bibitem{burattini2009comparative}
Laura Burattini, Silvia Bini, and Roberto Burattini.
\newblock Comparative analysis of methods for automatic detection and
  quantification of microvolt t-wave alternans.
\newblock {\em Medical engineering \& physics}, 31(10):1290--1298, 2009.

\bibitem{hannun2019cardiologist}
Hannun~Awni Y, Rajpurkar Pranav, Haghpanahi Masoumeh, Tison~Geoffrey H, Bourn
  Codie, Turakhia~Mintu P, and Ng~Andrew Y.
\newblock Cardiologist-level arrhythmia detection and classification in
  ambulatory electrocardiograms using a deep neural network.
\newblock {\em Nature medicine}, 25(1):65--69, 2019.

\bibitem{ptbxl}
Wagner P., Strodthoff N., and Bousseljot~RD. et~al.
\newblock Ptb-xl, a large publicly available electrocardiography dataset.
\newblock {\em Scientific Data}, 2020.

\bibitem{physionet}
AL~Goldberger, LA~Amaral, L~Glass, JM~Hausdorff, PC~Ivanov, RG~Mark, JE~Mietus,
  GB~Moody, CK~Peng, and HE~Stanley.
\newblock Physiobank, physiotoolkit, and physionet: components of a new
  research resource for complex physiologic signals.
\newblock {\em Circulation}, 101(23):E215—20, June 2000.

\bibitem{bank2021autoencoders}
Dor Bank, Noam Koenigstein, and Raja Giryes.
\newblock Autoencoders, 2021.

\bibitem{gu2022modeling}
Gu~Shawn, Jiang Meng, Guzzi~Pietro Hiram, and Tijana Milenkovi{\'c}.
\newblock Modeling multi-scale data via a network of networks.
\newblock {\em Bioinformatics}, 38(9):2544--2553, 2022.

\bibitem{dong2015imagesr}
Chao Dong, Chen~Change Loy, Kaiming He, and Xiaoou Tang.
\newblock Image super-resolution using deep convolutional networks, 2015.

\bibitem{lomoio2023ecg}
Lomoio Ugo, Vizza Patrizia, Giancotti Raffaele, Tradigo Giuseppe, Petrolo
  Salvatore, Flesca Sergio, Hiram~Guzzi Pietro, and Veltri Pierangelo.
\newblock Autan-ecg: An autoencoder based system for anomaly detection in ecg
  signals.
\newblock {\em "techrxiv"}, 2023.

\bibitem{zhang2018CAE}
Yifei Zhang.
\newblock A better autoencoder for image: Convolutional autoencoder.
\newblock 2018.

\bibitem{fariha2020ecgnoise}
Fariha Ziti, Ikeura Ryojun, Hayakawa Soichiro, and Tsutsumi Shigeyoshi.
\newblock An analysis of the effects of noisy electrocardiogram signal on
  heartbeat detection performance.
\newblock {\em Bioengineering}, 7:53, 06 2020.

\bibitem{perez2018artifacts}
Pérez-Riera AR, Barbosa-Barros R, Daminello-Raimundo R, and de~Abreu~LC.
\newblock Main artifacts in electrocardiography.
\newblock {\em Ann Noninvasive Electrocardiol.}, 2018.

\bibitem{dcae2008}
Vincent Pascal, Larochelle Hugo, Bengio Yoshua, and Manzagol Pierre-Antoine.
\newblock Extracting and composing robust features with denoising autoencoders.
\newblock page 1096–1103, 2008.

\bibitem{dcae2016images}
Gondara Lovedeep.
\newblock Medical image denoising using convolutional denoising autoencoders.
\newblock In {\em 2016 IEEE 16th International Conference on Data Mining
  Workshops (ICDMW)}, pages 241--246, 2016.

\bibitem{guzzi2022editorial}
Guzzi~Pietro Hiram and Zitnik Marinka.
\newblock Editorial deep learning and graph embeddings for network biology.
\newblock {\em IEEE/ACM Transactions on Computational Biology and
  Bioinformatics}, 19(02):653--654, 2022.

\bibitem{lio2020srmri}
Tan C, Zhu J, and Liò Pietro.
\newblock Arbitrary scale super-resolution for brain mri images. artificial
  intelligence applications and innovations.
\newblock {\em Artificial Intelligence Applications and Innovations}, 2020.

\bibitem{lio2019lesionsr}
Jin Zhu, Guang Yang, and Pietro Li.
\newblock How can we make gan perform better in single medical image
  super-resolution? a lesion focused multi-scale approach, 2019.

\bibitem{kuleshov2017audio}
Volodymyr Kuleshov, S.~Zayd Enam, and Stefano Ermon.
\newblock Audio super resolution using neural networks, 2017.

\bibitem{castiglioni2003interpolation}
Castiglioni P., Piccini L., and Di~Rienzo M.
\newblock Interpolation technique for extracting features from ecg signals
  sampled at low sampling rates.
\newblock In {\em Computers in Cardiology, 2003}, pages 481--484, 2003.

\bibitem{Han2013ComparisonOC}
Dianyuan Han.
\newblock Comparison of commonly used image interpolation methods.
\newblock 2013.

\bibitem{gpu2008}
Owens~John D., Houston Mike, Luebke David, Green Simon, Stone~John E., and
  Phillips~James C.
\newblock Gpu computing.
\newblock {\em Proceedings of the IEEE}, 96(5):879--899, 2008.

\bibitem{chen2022SRreview}
Honggang Chen, Xiaohai He, Linbo Qing, Yuanyuan Wu, Chao Ren, Ray~E. Sheriff,
  and Ce~Zhu.
\newblock Real-world single image super-resolution: A brief review.
\newblock {\em Information Fusion}, 79:124--145, 2022.

\bibitem{mekhlafi2024SRreview}
Al-Mekhlafi H. and Liu S.
\newblock Single image super-resolution: a comprehensive review and recent
  insight.
\newblock {\em Front. Comput. Sci}, 2024.

\bibitem{SRCNN}
Chao Dong, Chen~Change Loy, Kaiming He, and Xiaoou Tang.
\newblock Image super-resolution using deep convolutional networks, 2015.

\bibitem{FSRCNN}
Chao Dong, Chen~Change Loy, and Xiaoou Tang.
\newblock Accelerating the super-resolution convolutional neural network, 2016.

\bibitem{lyu2020progressivesr}
Lyu Q, Shan H, Steber C, Helis C, Whitlow C, Chan M, and Wang G.
\newblock Multi-contrast super-resolution mri through a progressive network.
\newblock {\em IEEE Trans Med Imaging}, 2020.

\bibitem{SRECG}
Tsai-Min Chen, Yuan-Hong Tsai, Huan-Hsin Tseng, Kai-Chun Liu, Jhih-Yu Chen,
  Chih-Han Huang, Guo-Yuan Li, Chun-Yen Shen, and Yu~Tsao.
\newblock Srecg: Ecg signal super-resolution framework for portable/wearable
  devices in cardiac arrhythmias classification.
\newblock {\em IEEE Transactions on Consumer Electronics}, 69(3):250--260,
  2023.

\bibitem{SRResNet}
Christian Ledig, Lucas Theis, Ferenc Huszar, Jose Caballero, Andrew~P. Aitken,
  Alykhan Tejani, Johannes Totz, Zehan Wang, and Wenzhe Shi.
\newblock Photo-realistic single image super-resolution using a generative
  adversarial network.
\newblock {\em CoRR}, abs/1609.04802, 2016.

\bibitem{CPSC2018}
F.~F. Liu, C.~Y. Liu*, L.~N. Zhao, X.~Y. Zhang, X.~L. Wu, X.~Y. Xu, Y.~L. Liu,
  C.~Y. Ma, S.~S. Wei, Z.~Q. He, J.~Q. Li, and N.~Y. Kwee.
\newblock An open access database for evaluating the algorithms of ecg rhythm
  and morphology abnormal detection.
\newblock {\em Journal of Medical Imaging and Health Informatics}, 2018.

\bibitem{cao2022referencebased}
Jiezhang Cao, Jingyun Liang, Kai Zhang, Yawei Li, Yulun Zhang, Wenguan Wang,
  and Luc~Van Gool.
\newblock Reference-based image super-resolution with deformable attention
  transformer, 2022.

\bibitem{SRCNN-ECG}
Christina~Perinbam Kaniraja, Vani~Devi M., and Deepak Mishra.
\newblock A deep learning framework for electrocardiogram (ecg) super
  resolution and arrhythmia classification.
\newblock {\em Res. Biomed. Eng.}, 2024.

\bibitem{resnet}
Kaiming He, Xiangyu Zhang, Shaoqing Ren, and Jian Sun.
\newblock Deep residual learning for image recognition, 2015.

\bibitem{scpecg}
Paul Rubel, Jocelyne Fayn, Peter Macfarlane, Danilo Pani, Alois Schlögl, and
  Alpo Värri.
\newblock The history and challenges of scp-ecg: The standard communication
  protocol for computer-assisted electrocardiography.
\newblock {\em Hearts}, 2:384--409, 08 2021.

\bibitem{aha1990standard}
J.~J. Bailey, A.~S. Berson, A.~Garson Jr, L.~G. Horan, P.~W. Macfarlane, D.~W.
  Mortara, and C.~Zywietz.
\newblock Recommendations for standardization and specifications in automated
  electrocardiography: bandwidth and digital signal processing. a report for
  health professionals by an ad hoc writing group of the committee on
  electrocardiography and cardiac electrophysiology of the council on clinical
  cardiology, american heart association.
\newblock {\em AHA Circulation}, 1990.

\bibitem{fuentes2012bpfilters}
Buendía-Fuentes F, Arnau-Vives MA, Arnau-Vives A, Jiménez-Jiménez Y,
  Rueda-Soriano J, Zorio-Grima E, Osa-Sáez A, Martínez-Dolz LV, Almenar-Bonet
  L, and Palencia-Pérez MA.
\newblock High-bandpass filters in electrocardiography: Source of error in the
  interpretation of the st segment.
\newblock {\em ISRN Cardiol}, 2012.

\bibitem{neurokit2}
Makowski D., Pham T., and Lau~Z.J. et~al.
\newblock Neurokit2: A python toolbox for neurophysiological signal processing.
\newblock {\em Behavior Research Methods}, 53:1689–1696, 2021.

\bibitem{mse}
Mark~D. Schluchter.
\newblock {\em Mean Square Error}.
\newblock John Wiley \& Sons, Ltd, 2014.

\bibitem{rmse}
T.~O. Hodson.
\newblock Root-mean-square error (rmse) or mean absolute error (mae): when to
  use them or not.
\newblock {\em Geoscientific Model Development}, 15(14):5481--5487, 2022.

\bibitem{ssim}
Zhou Wang, A.C. Bovik, H.R. Sheikh, and E.P. Simoncelli.
\newblock Image quality assessment: from error visibility to structural
  similarity.
\newblock {\em IEEE Transactions on Image Processing}, 13(4):600--612, 2004.

\bibitem{snr}
Tongtong Yuan, Weihong Deng, Jian Tang, Yinan Tang, and Binghui Chen.
\newblock Signal-to-noise ratio: A robust distance metric for deep metric
  learning, 2019.

\bibitem{psnr}
Alain Horé and Djemel Ziou.
\newblock Image quality metrics: Psnr vs. ssim.
\newblock In {\em 2010 20th International Conference on Pattern Recognition},
  pages 2366--2369, 2010.

\end{thebibliography}

\end{document}